\documentclass[10pt,a4paper,twocolumn]{article}
\usepackage[printonlyused,withpage]{acronym}

\usepackage[acronym]{glossaries}

\newacronym{gc}{GC}{Granger causality}
\newacronym{csl}{CSL}{causal structure learning}
\newacronym{cbn}{CBN}{causal Bayesian network}
\newacronym{cma}{CMA}{causal Markov assumption}
\newacronym{cfa}{CFA}{causal faithfulness assumption}
\newacronym{ml}{ML}{machine learning}
\newacronym{mec}{MEC}{Markov equivalence class}
\newacronym{dag}{DAG}{direct acyclic graph}
\newacronym{rccp}{RCCP}{Reichenbach's common cause principle}
\newacronym{mvgc}{MVGC}{Multivariate Granger causality}
\newacronym{bvgc}{BVGC}{Bivariate Granger causality}
\newacronym{cgc}{c-GC}{causalised Granger causality}
\newacronym{fcgc}{c-GC*}{fully causalised Granger causality}
\newacronym{iid}{iid}{independent and identically distributed}
\newacronym{mi}{MI}{mutual information}
\newacronym{fpr}{FPR}{false positive rate}
\newacronym{dpf}{dpf}{days post fertilization}

\usepackage{graphicx}
\usepackage{microtype}
\usepackage{cancel}
\usepackage{float}
\usepackage{url}
\usepackage{xcolor}
\usepackage{amsfonts}
\usepackage{amsmath}
\usepackage{hyperref}
\usepackage{tikz}
\usepackage{array}
\usepackage{caption}
\usepackage{subcaption}
\usepackage{longtable}
\usepackage{multirow}
\usetikzlibrary{cd}
\usepackage{bm}
\usepackage{color}
\usepackage[numbers,sort&compress]{natbib}
\usepackage{amsthm}
\usepackage{amssymb}
\usepackage{hyperref}
\usepackage[T1]{fontenc}
\usepackage{babel}
\usepackage[left=0.8in,right=0.8in,top=.8in,bottom=.8in]{geometry}

\newtheorem{remark}{Remark}
\usepackage[ruled]{algorithm2e}
\usepackage{algorithmic}
\usetikzlibrary{arrows}
\usepackage[normalem]{ulem}
\usepackage{authblk}

\makeatletter
\newcommand{\xleftrightarrow}[2][]{\ext@arrow 3359\leftrightarrowfill@{#1}{#2}}
\newcommand{\xdashrightarrow}[2][]{\ext@arrow 0359\rightarrowfill@@{#1}{#2}}
\newcommand{\xdashleftarrow}[2][]{\ext@arrow 3095\leftarrowfill@@{#1}{#2}}
\newcommand{\xdashleftrightarrow}[2][]{\ext@arrow 3359\leftrightarrowfill@@{#1}{#2}}
\def\rightarrowfill@@{\arrowfill@@\relax\relbar\rightarrow}
\def\leftarrowfill@@{\arrowfill@@\leftarrow\relbar\relax}
\def\leftrightarrowfill@@{\arrowfill@@\leftarrow\relbar\rightarrow}
\def\arrowfill@@#1#2#3#4{%
  $\m@th\thickmuskip0mu\medmuskip\thickmuskip\thinmuskip\thickmuskip
   \relax#4#1
   \xleaders\hbox{$#4#2$}\hfill
   #3$%
}

\definecolor{orcidlogocol}{HTML}{A6CE39}
\DeclareRobustCommand{\orcidicon}{\tikz[baseline=-0.6ex]\node[circle,fill=orcidlogocol,inner sep=0.3pt]{\color{white}\sffamily\bfseries\scriptsize iD};}
\DeclareRobustCommand{\orcidID}[1]{\,\href{#1}{\orcidicon}}

\title{\textbf{Conditioning-Depth Diagnostics for Hidden Memory in Temporal Causal Discovery}}
\author{Adedayo, S. A.\orcidID{https://orcid.org/0000-0003-2990-142X}\thanks{Correspondence concerning this article should be addressed to Adedayo, S. A.}}
\affil{UniVie Doctoral School of Computer Science, University of Vienna, Austria}
\date{}
\begin{document}
\twocolumn[
    \maketitle
    \begin{center}
    \rule{\textwidth}{2pt}
    \end{center}
    \begin{center}
    \begin{minipage}{0.87\textwidth}
    \begin{abstract}
    \sloppy
    Reliable causal discovery in time series requires conditioning sets that capture the system state. When predictive history is omitted, residual dependence can appear as direct causal links. We test state adequacy by measuring how inferred graphs change as conditioning depth increases while the reported causal lag stays fixed. Under an adequate finite-order Markov representation, graphs should stabilize once enough observed history is conditioned on; latent common drive, omitted lags, nonstationarity, and measurement dynamics can instead produce depth sensitivity. We formalize this idea with graph instability statistics and evaluate c-GC and c-GC*, the two learners whose depth parameter implements a matched fixed-horizon history intervention. PCMCI+ and JPCMCI+ are excluded from the primary comparison because adaptive parent selection makes nominal depth edge-specific. In paired simulations, a clean order-1 process was stable in every repeat, whereas an AR(1) latent common driver produced positive instability in all c-GC repeats and eight of ten c-GC* repeats. In calcium-imaging recordings, connectivity drops at the first transition beyond the one-lag baseline and then levels off, but B=200 bootstrap calibration does not reject the fitted order-1 null. The workflow therefore flags hidden memory or observed-state inadequacy without identifying the generating mechanism or recovering a latent graph.
    \end{abstract}
    \fussy

    \end{minipage}
    \end{center}
    \begin{center}
    \par\vspace{1em}\noindent
    \rule{\textwidth}{2pt}
    \end{center}
    \vspace{0.5cm}
]

    \section{Introduction}

Causal structure learning (CSL) from observational time series is difficult because predictive dependence need not be causal dependence \citep{granger1969,geweke1982linear,pearl1995diagrams,pearl2009causality,eichler2007granger,barnett2009granger,runge2019pcmci}. Autocorrelation, lagged interactions, contemporaneous dependence, and unobserved common drivers can all create associations among observed variables \citep{eichler2007granger,eichler2012graphical,eichler2013multiple,runge2019pcmci,runge2020pcmcip,runge2023review,chen2024discovery}. The risk is largest when hidden processes affect several observed variables: edges then appear direct only because the algorithm conditioned on an incomplete state \citep{spirtes1995fci,spirtes2000causation,richardson2002ancestral,zhang2008orientation,entner2010fci,malinsky2018timeseries,gerhardus2020lpcmci}.

Temporal CSL methods reduce this risk by conditioning on lagged histories. Granger-style methods ask whether a source adds predictive information after relevant past variables are included, while PCMCI variants refine the conditioning set with lag-aware parent selection and conditional-independence testing \citep{granger1969,geweke1982linear,geweke1984conditional,eichler2007granger,barnett2009granger,runge2018reconstruction,runge2019pcmci,runge2020pcmcip}. These methods still require a state representation. If the chosen lag order is too shallow, sampling hides intermediate dynamics, or hidden processes carry memory across time, graph estimates can change with conditioning depth \citep{eichler2013multiple,hyttinen2017subsampled,runge2018reconstruction,malinsky2018timeseries,runge2020pcmcip,runge2023review}.

We turn this failure mode into a diagnostic. Instead of treating one learned graph as final, we rerun the learner across conditioning depths and ask whether the inferred graph stabilizes. The primary experiments use causalised Granger causality (\acrshort{cgc}) and its extension (\acrshort{fcgc}) because both keep the reported graph horizon fixed while changing observed history depth \citep{adedayo2025re}. Tigramite PCMCI+ and JPCMCI+ remain important temporal methods \citep{runge2020pcmcip,gunther2023jpcmciplus}, but their adaptive searches do not implement the same fixed-horizon intervention. LPCMCI is therefore treated separately as a latent-variable-aware sensitivity analysis \citep{gerhardus2020lpcmci}.

The diagnostic target is hidden memory, not unique latent-confounder identification. Latent common drive is one source of hidden memory; omitted lags, nonstationarity, measurement error, undersampling, nonlinear dynamics, and finite-sample testing error are others \citep{hyttinen2017subsampled,runge2018reconstruction,runge2020pcmcip,gerhardus2020lpcmci,runge2023review}. The diagnostic asks whether the selected observed state is stable enough for causal interpretation. Attributing instability to a specific mechanism requires additional graphical, algebraic, non-Gaussian, proxy, environmental, measurement, or parametric assumptions \citep{spirtes1995fci,richardson2002ancestral,tian2002testable,colombo2012rfci,ogarrio2016gfci,zhang2008orientation,shimizu2006lingam,hoyer2008hidden,karlsson2023detecting,miao2018proxy,tchetgen2020proximal,chen2024discovery}.

This manuscript contributes a conditioning-depth model check for temporal CSL. We formalize graph stability as a Markovianity diagnostic, define graph instability statistics, and use VAR and residual-bootstrap calibration to attach null reference distributions to real-data curves \citep{kunsch1989,politis1994}. The analyses include automatic depth selection, FDR-controlled edgewise localization, controls, mechanism ablations, sample-size and power analysis, and nonlinear-CI sensitivity checks \citep{benjamini1995}. Synthetic experiments test the signature with known ground truth, and calcium-imaging v2a-RSN data are reported with both descriptive depth trajectories and B=200 bootstrap-calibrated non-rejection under a fitted order-1 null \citep{yu2009gpfa,paninski2010state,ahrens2013whole}. Sections~\ref{sec:previous-work}--\ref{sec:conclusions} review related work, define the method, present the experiment, and summarize limitations.

    \section{Previous Work} \label{sec:previous-work}

Latent confounding limits causal claims from observational data \citep{pearl1995diagrams,pearl2009causality,spirtes1995fci,spirtes2000causation,richardson2002ancestral,miao2018proxy}. A latent confounder is an unobserved cause of two or more observed variables; in time series, autocorrelation, lagged effects, contemporaneous dependence, and measurement dynamics can create similar apparent memory \citep{eichler2007granger,eichler2010graphical,eichler2012graphical,eichler2013multiple,runge2018reconstruction,assaad2022survey,runge2023review}. Testability is the main obstacle. One observational distribution can fit several hidden-variable explanations; methods therefore need graph constraints, algebraic restrictions, non-Gaussianity, proxies, environments, temporal order, negative controls, or parametric models \citep{verma1990equivalence,spirtes1995fci,richardson2002ancestral,ali2009markov,zhang2008orientation,tian2002testable,shimizu2006lingam,hoyer2008hidden,miao2018proxy,karlsson2023detecting,gerhardus2020lpcmci,glymour2019review,sadeghi2022conditions}.

\subsection{Scope and Background}

Let $X=\{X_1,\ldots,X_d\}$ be observed variables and $C=\{C_1,\ldots,C_q\}$ hidden variables. In an SCM,
\begin{equation}
  X_i = f_i(\mathrm{Pa}(X_i), U_i),
\end{equation}
with independent exogenous noises unless common causes represent dependence \citep{pearl1995diagrams,pearl2009causality,spirtes2000causation,peters2017elements}. The analyst observes only
\begin{equation}
  P_X(x)=\int P_{X,C}(x,c)\,dc .
\end{equation}
The question is whether $P_X$ violates every causally sufficient model in the chosen class.

Latent projection maps a DAG over $X\cup C$ to an ADMG: directed edges summarize hidden-node directed paths, and bidirected edges summarize latent common causes \citep{richardson2002ancestral,pearl2009causality,shpitser2006id,shpitser2008hierarchy,richardson2023nested}. In ADMGs and maximal ancestral graphs, m-separation replaces d-separation; PAGs represent equivalence classes when latent or selection variables may exist \citep{spirtes1995fci,spirtes2000causation,colombo2012rfci,ali2009markov,zhang2008orientation}. Thus representing hidden variables, detecting causal-sufficiency violations, and estimating effects are distinct tasks.

\subsection{Method Categories}

Three method families matter here. Graphical methods represent hidden variables with mixed graphs. Model-specific methods add assumptions for detection or adjustment. Time-series methods use temporal order to test state adequacy. We follow the third route: state adequacy, not recovery of a unique latent graph.

\subsubsection{Latent Graphical Methods}

FCI is the classical constraint-based method without causal sufficiency \citep{spirtes1995fci,spirtes2000causation}. It removes adjacencies with conditional-independence tests and returns a PAG; bidirected or partial marks show hidden variables are compatible with the independencies, but rarely identify them. RFCI lowers the test burden \citep{colombo2012rfci}, GFCI adds score-based search \citep{ogarrio2016gfci}, and \citet{zhang2008orientation} clarifies orientation. Markov equivalence explains why PAGs can remain broad even in population \citep{ali2009markov}. PC search shares the sensitivity to sparsity, test calibration, order, and faithfulness \citep{kalisch2007pc,colombo2014order,sadeghi2022conditions,glymour2019review}.

Mixed-graph theory also separates representation from identification. DAGs and \textit{do}-calculus identify effects when a graph or equivalence class is known \citep{pearl1995diagrams,pearl2009causality}. Latent projections, Verma constraints, and nested Markov models describe hidden-variable implications beyond ordinary conditional independence \citep{richardson2002ancestral,shpitser2006id,shpitser2008hierarchy,richardson2023nested}; they do not recover the hidden variables.

\subsubsection{Model-Based Methods}

Algebraic methods use Verma constraints, tetrads, rank restrictions, cumulants, or trek-separation relations \citep{verma1990equivalence,tian2002testable,silva2006latent,kummerfeld2016fofc,sullivant2010trek,foygel2012halftrek,chen2024cumulants}. They work under measurement, linearity, or non-Gaussian assumptions, but are less general for nonlinear time series. Linear non-Gaussian models, including LiNGAM variants, use residual independence for identification \citep{shimizu2006lingam,hoyer2008hidden,shimizu2011directlingam,maeda2020rcd,tramontano2024lingam}.

Other approaches add auxiliary structure. Proxy and negative-control methods need exclusion, relevance, bridge, or completeness assumptions \citep{lipsitch2010negative,kuroki2014proxy,miao2018proxy,shi2020negative,shi2020multiply,tchetgen2024proximal,cui2023semiparametric,penning2023negative}. Deep latent-variable and deconfounder methods infer substitute confounders from latent representations or multiple causes \citep{louizos2017causal,wang2019deconfounder,damour2019multicause,bica2020time,chen2024discovery}. Multiple-environment methods use invariant prediction or mechanism variation when the environment assumptions hold \citep{karlsson2023detecting,peters2016invariant,pfister2018invariant}.

\subsubsection{Time-Series Discovery}

In Granger frameworks, $X^j$ does not cause $X^i$ when the past of $X^j$ adds no predictive information for $X^i_t$ after conditioning on relevant observed history \citep{granger1969,geweke1982linear,eichler2007granger,barnett2009granger}. Graphical time-series and VAR models state the same idea through lagged conditional independence and dynamic mixed graphs \citep{eichler2010graphical,eichler2012graphical,eichler2013multiple}. Classical lag-order and residual-diagnostic checks are therefore complementary to the present graph-stability diagnostic: they assess whether a fitted time-series null is adequate, while the proposed statistic asks whether the learned graph changes when more observed history is conditioned on. In neural data, state-space modeling also addresses filtered or noisy observations before causal interpretation \citep{paninski2010state}. Surveys cover Granger, constraint-based, score-based, noise-based, topology-based, and hybrid methods \citep{assaad2022survey,runge2023review}.

\acrshort{cgc} and \acrshort{fcgc} place Granger-style tests in causal Bayesian network language and Reichenbach semantics \citep{adedayo2025re}. They retain an edge only when bivariate and conditional tests support the same directed relation, and their conditioning depth is explicit. PCMCI and PCMCI+ use parent selection and momentary conditional independence \citep{runge2018reconstruction,runge2019pcmci,runge2020pcmcip}; JPCMCI+ adds multiple datasets and context variables \citep{gunther2023jpcmciplus}. SVAR-FCI and temporal FCI adapt latent-variable discovery to stationary time series \citep{entner2010fci,malinsky2018timeseries}; LPCMCI represents hidden common causes through PAG marks \citep{gerhardus2020lpcmci,runge2023review}. Graphical Gaussian work and direct time-series detectors show how hidden processes induce spurious Granger relations \citep{eichler2010graphical,liu2023strength}. Neural and invariance-based Granger methods handle nonlinear dynamics or heterogeneous environments \citep{tank2022neural,zhou2024jrngc,zhang2025invargc}.

\subsection{Positioning}

This paper is a hidden-memory model check. If an observed process is well represented by a finite-order Markov state, deeper observed history should not systematically change the population constraint-based graph. Persistent depth sensitivity means the selected state omits predictive memory. Latent confounding can produce that signal, but so can omitted lags, undersampling, nonstationarity, measurement error, nonlinear dynamics, and finite-sample conditional-independence error \citep{hyttinen2017subsampled,runge2018reconstruction,gerhardus2020lpcmci,runge2023review}.

    \section{Method} \label{sec:method}

\subsection{Setup and Notation}

Let $X_t = (X_t^{(1)}, \dots, X_t^{(d)})$ denote the observed time series and let $C_t$ denote an unobserved process that may influence several observed variables. Hidden memory means predictive state information absent from the selected observed-state representation. A persistent latent common driver is one mechanism, but omitted lags, temporal aggregation, nonstationarity, measurement dynamics, and other unmodeled dependence can produce the same operational problem.

The data are assumed to arise from a dynamic causal system with finite but unknown effective memory. Let $p$ be conditioning depth, the number of past observed states included in the conditioning set. The implementation uses $n_{\mathrm{pasts}}$ for the same quantity:
\[
  p \equiv n_{\mathrm{pasts}}.
\]
The reported graph horizon is controlled separately. In one-lag \acrshort{cgc} analyses, \(p=1\) is the baseline graph horizon and \(p=2\) is the first extra-history diagnostic condition. We run the grid from \(p=1\) because \(\hat{A}^{(1)}\) is needed to compute the first adjacent-depth change.

For a chosen depth, the learner is applied to
\(\mathcal{X}^{(p)}_t = \{X_{t-1}, X_{t-2}, \dots, X_{t-p}\},\)
producing an inferred graph $\hat{G}^{(p)}$ or, in simulations, recovery metrics against the known graph. The diagnostic question is whether the graph stabilizes once enough observed history is included.

\subsection{State-Adequacy Principle}

If the observed process is adequately represented by a finite-order Markov state, deeper lags should add no systematic information after the relevant history has been conditioned on. Shallow conditioning can instead leave open dependence paths, creating spurious links that disappear or change as \(p\) increases. This depth sensitivity diagnoses hidden memory relative to the selected state representation; it does not identify the mechanism.

\begin{figure}
\centering
    \begin{subfigure}[b]{0.24\textwidth}
    \centering
        \includegraphics[width=\linewidth]{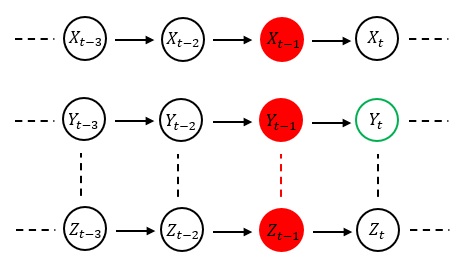}
    \caption{}
    \label{fig:nonConfCase}
    \end{subfigure}
    \hfill
    \begin{subfigure}[b]{0.24\textwidth}
    \centering
        \includegraphics[width=\linewidth]{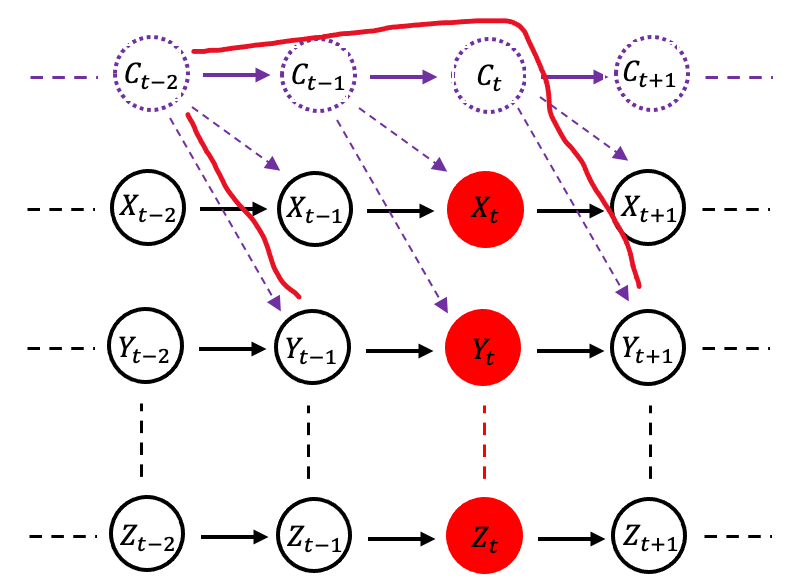}
    \caption{}
    \label{fig:confCase}
    \end{subfigure}
\caption{Schematic intuition for the Markovianity diagnostic. Fig.~\ref{fig:nonConfCase} shows an adequate observed Markov state; Fig.~\ref{fig:confCase} shows hidden temporal dependence carried outside the conditioned state.}
\label{fig:MarkTest}
\end{figure}

Figure~\ref{fig:MarkTest} gives the intuition. In Fig.~\ref{fig:nonConfCase}, conditioning on the intervening observed slice blocks deeper observed history. In Fig.~\ref{fig:confCase}, the latent process creates a route outside that conditioned slice, so earlier activity can remain predictive even after the observed variables at the previous time are included.

\paragraph{Heuristic principle.}
Assume that, after conditioning on observed parents up to lag $\tau$, deeper observed lags contain no additional information for the population-level conditional tests used by the selected learner. Then, for $p \geq \tau$, increasing conditioning depth from $p$ to $p+1$ should not systematically change the population graph. This is a model-checking principle rather than a formal monotonicity guarantee for finite-sample \acrshort{cgc} or \acrshort{fcgc}: adding superfluous variables can still change tests through finite-sample power, collinearity, or implementation details.

\begin{remark}
Persistent changes in $\hat{G}^{(p)}$ diagnose hidden memory relative to the chosen observed state. Latent common drive, omitted lag order, temporal aggregation, nonstationarity, and observation dynamics can all produce this warning signal.
\end{remark}

\subsection{Assumptions}

The interpretation requires:
\begin{itemize}
    \item \acrlong{cma}: a variable is independent of non-effects given its direct causes;
    \item a depth grid that reaches the dominant observed lag order, \(p \geq \tau_{\max}\);
    \item hidden-memory mechanisms that transmit predictive information not summarized by the shallow observed state.
\end{itemize}
The contrast is
\[
\mathcal{H}_0: \; \parbox[t]{0.8\linewidth}{the observed process is adequately represented as Markov at the chosen order,}
\]
\[
\mathcal{H}_1: \; \parbox[t]{0.8\linewidth}{the selected observed-state representation omits predictive memory.}
\]

\subsection{Structure-Learning Embedding}

For depths $p \in \{p_{\min}, \dots, p_{\max}\}$, we run the selected learner repeatedly and record
\(\hat{G}^{(p_{\min})}, \hat{G}^{(p_{\min}+1)}, \dots, \hat{G}^{(p_{\max})}.\) In simulations, known ground truth allows direct recovery metrics: accuracy, precision, recall, \acrlong{fpr}, balanced accuracy, and F1. In observational data, no true graph is available, so we summarize graph stability through edge counts and adjacent-depth adjacency changes. These graph summaries are therefore a real-data diagnostic layer, not the primary simulation endpoint.

\subsection{Experimental Learners}

\acrshort{cgc} and \acrshort{fcgc} share a two-stage template. For each ordered source--target pair and depth $p$, the series is stacked into current variables and $p$ past observed states. The algorithm performs an unconditional source-target screen and then a conditional residual test, using circular-shift permutation tests based on correlation. An edge is retained only when both stages are significant at thresholds $\alpha$ and $\beta$.

The variants differ in conditioning sets. \acrshort{cgc} excludes the candidate source row, target row, and source history needed to avoid conditioning away the source signal. \acrshort{fcgc} uses the fully shifted observed history except candidate source and target rows, making it more conservative when true lags are heterogeneous. In the fixed-horizon runs, the tested causal horizon is kept at one lag. Increasing \(p\) only adds older observed states to the conditioning set; it does not change the reported edge horizon. Lag-specific detections are then collapsed to binary directed adjacencies, so \(\hat{A}^{(p)}_{ij}=1\) means that at least one retained fixed-horizon detection supports \(j\to i\). This adjacency-level collapse matches the diagnostic question--whether the reported directed graph changes with added history--but it does not resolve which specific lag caused the change.

\paragraph{Comparator eligibility and exclusion.}
The primary depth sweep requires a learner to hold the reported causal lag fixed while $p$ directly controls added observed history. \acrshort{cgc} and \acrshort{fcgc} meet this requirement. PCMCI+ does not expose an equivalent single control \citep{runge2020pcmcip}: changing its maximum lag changes both conditioning and the candidate effect horizon, while adapting \texttt{max\_conds} parameters changes PC1 search depth, retained parents, and MCI conditioning in edge-specific ways. JPCMCI+ inherits this mismatch and is designed for joint discovery across multiple datasets or contexts \citep{gunther2023jpcmciplus}. LPCMCI is retained as a separate latent-variable-aware sensitivity analysis, preserving its marked graph output, but it is not part of the matched $p$-indexed trajectory \citep{gerhardus2020lpcmci}. SVAR-FCI remains outside the present matched-depth analysis.

\subsection{Instability Calibration}

For observational data, let \(m=d(d-1)\) be the number of directed off-diagonal edge positions. For $p>p_{\min}$, define
\[
D_p = \frac{1}{m}\left\|\hat{A}^{(p)} - \hat{A}^{(p-1)}\right\|_1,
\] 
the fraction of directed edges whose status changes after adding one more lag. We also decompose changes into deletions and additions:
\[
D_p^- = \frac{1}{m}\sum_{i \neq j}\mathbf{1}\{\hat{A}^{(p-1)}_{ij}=1,\hat{A}^{(p)}_{ij}=0\},
\]
\[
D_p^+ = \frac{1}{m}\sum_{i \neq j}\mathbf{1}\{\hat{A}^{(p-1)}_{ij}=0,\hat{A}^{(p)}_{ij}=1\}.
\]
Early deletion-dominated changes are expected when shallow conditioning creates spurious links, but the method does not require that sign pattern.

The global summaries are:
\[
T_{\mathrm{obs}} = \max_{p=p_{\min}+1,\dots,p_{\max}} D_p
\], 
the maximum adjacent-depth change.
\[
S_{\mathrm{obs}} = \sum_{p=p_{\min}+1}^{p_{\max}} D_p
\], 
the cumulative instability. For one-lag real-data runs, \(D_2\) is the first diagnostic transition because it compares the one-lag baseline graph with the graph conditioned on one additional past state.

Calibration compares the observed curve with surrogate data generated under a fitted Markovian null. Choose a baseline order $p_0$ and fit a null model $\widehat{\mathcal{M}}_{p_0}$, implemented here as a VAR($p_0$) with a stacked lag design. Surrogates are generated either by Gaussian VAR simulation, iid residual bootstrap, or moving-block residual bootstrap; the last preserves short-range dependence under weak stationarity \citep{kunsch1989, politis1994}. The v2a analysis uses a VAR(1) moving-block residual-bootstrap null with block length 50 and $B=200$ replicates. The adequacy of this null is checked separately with information criteria, stability, multivariate residual whiteness, and univariate Ljung--Box diagnostics; these checks assess the calibration model rather than the graph-instability statistic itself.

For each surrogate $b$, the full depth sweep is rerun and
\[
T^{*(b)} = \max_{p=p_0+1, \dots, p_{\max}} D_p^{*(b)}
\]
is computed. The calibrated global $p$-value is
\[
\hat{p}_{\mathrm{global}} = \frac{1 + \sum_{b=1}^B \mathbf{1}\{T^{*(b)} \ge T_{\mathrm{obs}}(p_0)\}}{B+1},
\]
with an equivalent critical-value rule
\[
c_{1-\alpha} = \mathrm{Quantile}_{1-\alpha}\left(T^{*(1)}, \dots, T^{*(B)}\right).
\]
Pointwise bands use the corresponding quantiles of \(D_p^{*(b)}\), and a first-break depth can be reported when the observed curve first exceeds the pointwise band.

Edgewise localization is a separate layer. For each directed edge, the implementation records observed additions, deletions, first appearance or deletion depth, instability frequency, bootstrap null frequency, an empirical bootstrap $p$-value, and a Benjamini--Hochberg adjusted $q$-value \citep{benjamini1995}. These quantities identify which links drive graph-level changes descriptively; they become calibrated discoveries only after multiplicity correction.

\subsection{Diagnostics and Controls}

Automatic depth-selection rules summarize $\{D_p\}$ into a recommended depth $p_\star$ when the curve stabilizes. The rules include threshold-based stabilization, relative-drop stabilization, and a bootstrap-band rule that selects the first depth after which the curve remains inside the null band. If no stable depth is found, the analysis reports no stable depth.

Controls and diagnostics are reported as analysis layers. Negative controls use known Markovian or shuffled data; positive controls introduce hidden memory through latent common drive, hidden nodes, omitted lag order, observation artifacts, or nonstationarity. Mechanism ablations group these sources into hidden-state, observation-artifact, and nonstationarity families. Sample-size and power analyses vary sample size, dimensionality, edge density, confounder strength, latent autocorrelation, noise scale, repeats, and depth grid.

\subsection{Interpretation} \label{sec:inter_rule}

The diagnostic is operational:
\begin{enumerate}
    \item choose a baseline depth \(p_{\min}\) at least as large as the nominal graph horizon;
    \item run the selected learner over increasing \(p\);
    \item track recovery metrics in simulations and graph-stability summaries in observational data;
    \item treat stable trajectories as support for approximate Markovianity at the tested order;
    \item treat persistent depth sensitivity as evidence that the selected observed state omits predictive memory.
\end{enumerate}
When surrogate calibration is available, the same curves are supplemented by bootstrap bands, a global $p$-value for \(T_{\mathrm{obs}}\), and multiplicity-corrected edgewise localization. In the experiments below, the primary evidence is the conditioning-depth trajectory, with v2a calcium recordings additionally tested against the fitted B=200 moving-block bootstrap null.

    \section{Experiment} \label{sec:experiment}

We evaluate the Markovianity diagnostic in simulations with known graphs and in calcium imaging recordings without causal ground truth. The simulations test whether recovery metrics remain stable for Markovian data and shift when hidden memory is introduced. The calcium analysis tests whether inferred connectivity is stable as conditioning depth increases. Code and notebooks are available at \url{https://github.com/adedayoas91/hidden-confounding-diagnostics}.

With ground truth, we report accuracy, precision, recall, false positive rate (FPR), balanced accuracy, and F1 score as functions of conditioning depth. However, false-positive-sensitive metrics are the main diagnostic readout because deeper conditioning should prune spurious links when shallow states are incomplete. Without ground truth, we report edge counts, $D_p$, deletion and addition components, and $T_{\mathrm{obs}}$.

\subsection{Synthetic Validation}

Synthetic data follow the autoregressive setup in \cite{adedayo2025re}. Markovian systems use known lag-1 or lag-2 dynamics, and non-Markovian variants add a smooth unobserved driver. We evaluate aggregate directed adjacency recovery rather than exact lag assignment. Each scenario has 10 repeats. For both \acrshort{cgc} and \acrshort{fcgc}, $n_{\mathrm{perm}}=1000$, $\alpha=0.01$, and $\beta=0.001$. Single-lag figures sweep $n_{\mathrm{pasts}}=1,\ldots,7$ while multiple-lag figures sweep $n_{\mathrm{pasts}}=2,\ldots,7$. PCMCI-family exploratory runs are excluded by the eligibility rule in Section~\ref{sec:method}.

\subsubsection{Simulation Scenarios}

We specify the four graph instability scenarios directly as data-generating processes. Each uses \(d=10\), \(T=2000\), ten paired seeds, and a depth grid \(p=1,\ldots,7\), while the reported graph horizon stays fixed at one lag. Innovations are independent Gaussian noise terms.

\paragraph{(a) Order-1 unconfounded.}
The negative control is \(X_t=A_1X_{t-1}+\epsilon_t,\qquad \epsilon_t\sim\mathcal{N}(0,I),\)
where \(A_1\) is sparse and stable. The observed state is Markov of order one, so increasing conditioning depth should not change the fixed horizon graph. All ten paired repeats passed numerical quality control.

\paragraph{(b) Order-3 omitted history.}
The first omitted-history control is \(X_t=A_1X_{t-1}+A_2X_{t-2}+A_3X_{t-3}+\epsilon_t .\)
The learner still reports a one-lag graph, so the lag-2 and lag-3 terms are observed predictive memory left outside the shallow state. This family was excluded after three of ten repeats became numerically explosive.

\paragraph{(c) Latent common driver.}
The latent-driver control is \(H_t=0.8H_{t-1}+\eta_t,\qquad X_t=A_1X_{t-1}+wH_{t-1}+\epsilon_t .\)
The unobserved AR(1) process drives several observed coordinates through \(w\). After marginalizing \(H_t\), the observed process contains persistent hidden memory and common-cause dependence. All ten paired repeats passed numerical quality control.

\paragraph{(d) Variable-lag omitted history.}
The heterogeneous omitted-history control uses \(X_t=\sum_{\ell=1}^{3}A_{\ell}^{\mathrm{var}}X_{t-\ell}+\epsilon_t,\)
with lower-density lag matrices so different observed pairs can act at different delays. It tests whether depth sensitivity appears when the omitted memory is observed but distributed across lags. This family was excluded after three of ten repeats became numerically explosive.


\subsubsection{Graph Instability}

\begin{figure}[!htbp]
    \centering
    \includegraphics[width=.95\linewidth]{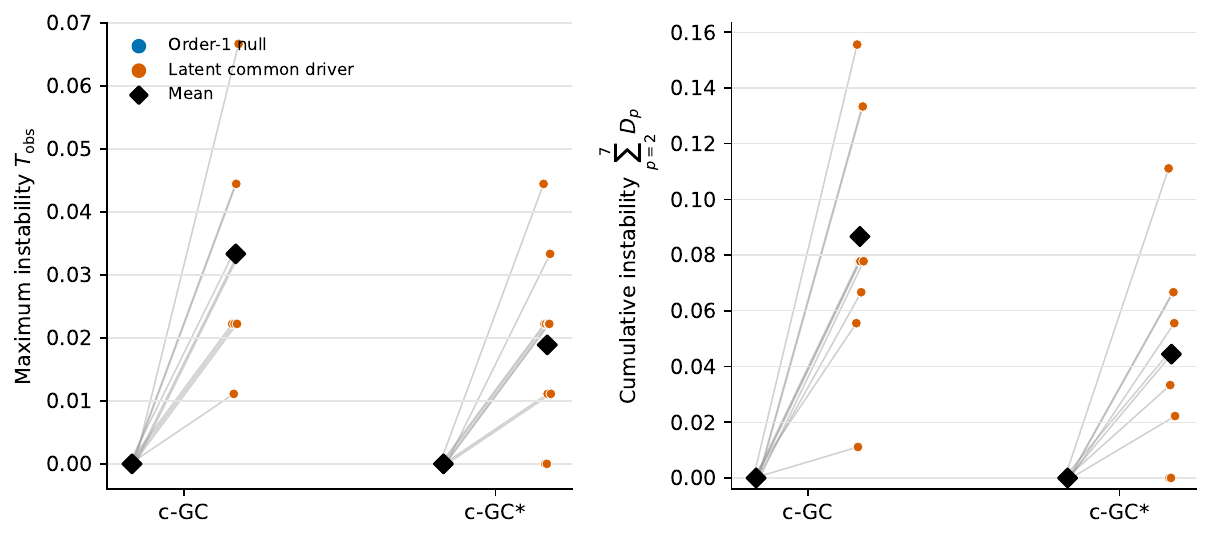}
    \caption{Graph instability summary for the numerically valid families. Points are paired repeats under the clean order-1 process and the AR(1) latent-common-driver process; grey lines join equal-seed pairs and diamonds show means. The clean graph is invariant, whereas the latent driver produces nonzero maximum and cumulative instability.}
    \label{fig:simulation-summary}
\end{figure}

The broad single- and multiple-lag recovery curves are supporting checks rather than the main result. They show the expected pattern: Markovian systems are mostly stable, hidden-memory perturbations improve false-positive-sensitive metrics as depth increases, and heterogeneous multi-lag systems are harder for the shallower learner (Supplementary Figure~\ref{fig:supp_recovery_curves}). The manuscript-facing statistic is graph instability itself. The clean order-1 process produced invariant graph trajectories across the depth grid for both methods. Adding an autocorrelated latent common driver moved the paired traces away from zero, with the stronger response under \acrshort{cgc}; corrected paired tests separated the latent-driver condition from the clean control (Figure~\ref{fig:simulation-summary}).

The order-3 and variable-lag families were not interpreted because numerical quality control found explosive repeats in each family. Both families were excluded wholesale rather than selectively retaining favorable repeats. Thus the retained simulation result supports detection of hidden memory from latent common drive, while broader mechanism generalization remains a supporting, not primary, claim.

\subsection{Calcium Imaging Data}

We next apply the diagnostic to four calcium-imaging recordings of v2a reticulospinal neurons (v2a-RSNs) in larval zebrafish (\textit{D. rerio}) \cite{carbo2023mesencephalic}. The calibrated analysis uses 25 traces from each recording, sampled as nine emitters and 16 receivers. This profile keeps the depth sweep and bootstrap calibration tractable while preserving both anatomical groups.

Both variants show deletion-dominated graph changes after the one-lag baseline. For \acrshort{cgc}, the first transition removes 17--20 more edges than it adds; for \acrshort{fcgc}, the corresponding net losses are 13--25 edges. The fully conditioned variant returns slightly fewer edges, but the qualitative profile is similar. The early drop is the biologically relevant warning: shallow one-lag graphs contain edges that disappear once one additional past state is admitted. Later transitions are smaller, so the descriptive profile suggests rapid stabilization rather than steadily increasing instability.

\begin{figure}[!htbp]
    \centering
    \includegraphics[width=.88\linewidth]{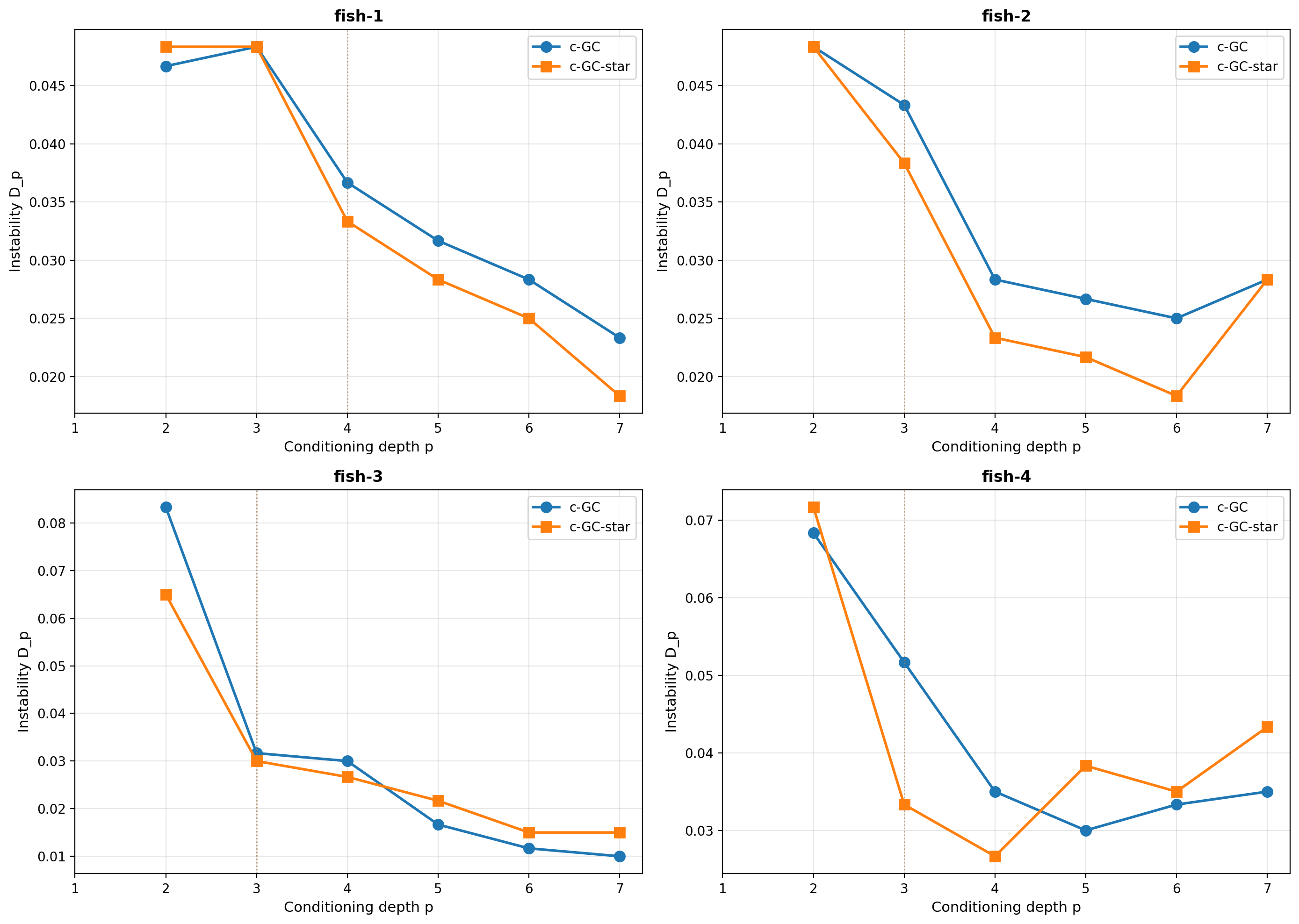}
    \caption{Conditioning-depth selection for the v2a-RSN profile with 25 traces: nine emitters and 16 receivers per recording. Each panel shows one recording, with observed adjacent-depth instability curves for \acrshort{cgc} and \acrshort{fcgc} plotted against the matched B=200 moving-block bootstrap null bands. The observed curves decrease after the first extra-history transition and stay below the calibrated critical envelope, motivating descriptive caution rather than formal rejection of the fitted order-1 null.}
    \label{fig:v2aMarkRes}
\end{figure}

\begin{figure}[!htbp]
    \centering
    \includegraphics[width=.95\linewidth]{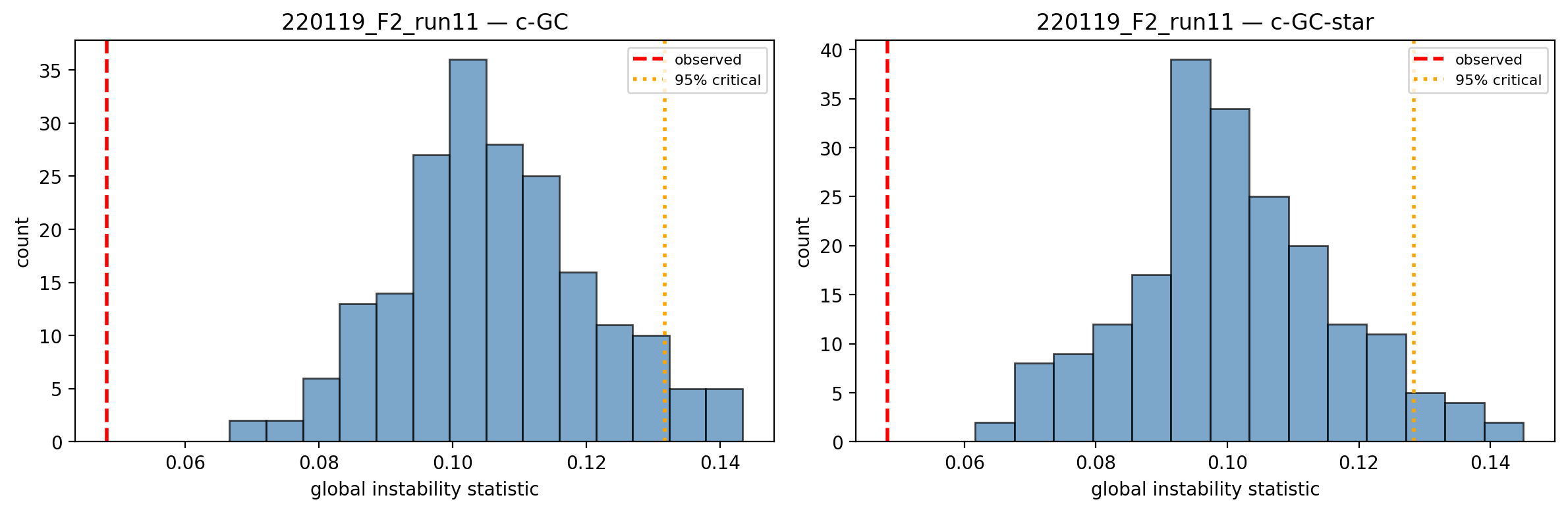}
    \caption{Representative bootstrap calibration for fish-1. The observed global statistic lies inside the \(B=200\) moving-block bootstrap null distribution, explaining why the calibrated analysis reports non-rejection despite descriptive graph changes.}
    \label{fig:v2a-bootstrap-example}
\end{figure}

The calibrated interpretation is conservative. All eight observed $T_{\mathrm{obs}}$ values are below their 95\% moving-block bootstrap critical values, so no recording-method pair rejects the fitted order-1 null through the graph-instability statistic. The real data show descriptive depth sensitivity, but the bootstrap global statistic does not provide statistically significant evidence that the fitted order-1 Markov model is inadequate. A separate VAR-null diagnostic gives useful context: the fitted VAR(1) models are stable, but residual whiteness tests reject in all four recordings, and information criteria usually prefer orders above one (Supplementary Table~\ref{tab:supp_var_null_diagnostics}). Thus the bootstrap non-rejection is conditional on an imperfect but stable calibration null. The fish-1 calibration shows the observed statistic inside the surrogate null distribution, illustrating why descriptive graph changes do not become a calibrated rejection (Figure~\ref{fig:v2a-bootstrap-example}). Edgewise localization is similarly conservative: across 1,207 calibrated edge rows, each compared against 200 bootstrap adjacency replicates, no edge survives false-discovery-rate correction at $q\leq0.05$. The calcium data therefore warn that shallow graphs are depth-sensitive, but do not provide calibrated evidence for a specific latent neural driver or localized edge-level discovery.

\begin{table}[!htbp]
\centering
\scriptsize
\setlength{\tabcolsep}{2pt}
\renewcommand{\arraystretch}{1.08}
\caption{Descriptive and calibrated v2a-RSN graph summaries for 25 traces per recording, combining nine emitters and 16 receivers. \(E_1\) and \(E_7\) are edge counts at the one-lag baseline and final depth; loss is the net edge loss at \(p=2\).}
\label{tab:v2a-summary}
\resizebox{\columnwidth}{!}{%
\begin{tabular}{@{}llrrlrrr@{}}
\hline
Method & Fish & \(E_1\) & \(E_7\) & Loss & \(T_{\mathrm{obs}}\) & \(c_{.95}\) & \(p\) \\
\hline
\multirow{4}{*}{\acrshort{cgc}}  & fish-1 & 96 & 47 & 20 (20.8\%) & 0.048 & 0.132 & 1.000 \\
 & fish-2 & 126 & 92 & 17 (13.5\%) & 0.048 & 0.127 & 1.000 \\
 & fish-3 & 133 & 101 & 20 (15.0\%) & 0.083 & 0.175 & 1.000 \\
 & fish-4 & 168 & 142 & 17 (10.1\%) & 0.068 & 0.148 & 1.000 \\
\hline
\multirow{4}{*}{\acrshort{fcgc}} & fish-1 & 94 & 43 & 25 (26.6\%) & 0.048 & 0.128 & 1.000 \\
 & fish-2 & 119 & 92 & 13 (10.9\%) & 0.048 & 0.117 & 1.000 \\
 & fish-3 & 122 & 100 & 13 (10.7\%) & 0.065 & 0.153 & 1.000 \\
 & fish-4 & 159 & 146 & 13 (8.2\%) & 0.072 & 0.135 & 0.990 \\
\hline
\end{tabular}
}
\end{table}

    \section{Conclusions}
\label{sec:conclusions}

This study recasts conditioning-depth sweeps as hidden-memory diagnostics for constraint-based temporal CSL. Stable trajectories support the selected observed Markov representation; persistent depth sensitivity indicates predictive memory missing from that representation. The method does not identify the mechanism or recover a latent graph.

The experiment supports this premise within the \acrshort{cgc} method family. In simulations, Markovian controls are stable while hidden-memory perturbations produce depth-dependent false-positive pruning. In the graph instability analysis, an order-1 control remains invariant, whereas an AR(1) latent common driver produces nonzero instability under both \acrshort{cgc} variants. The excluded higher-order families limit breadth across mechanisms, but not the narrower latent-driver conclusion. In calcium-imaging data, the 25-trace v2a-RSN profile combines nine emitters and 16 receivers and shows deletion-dominated first-depth changes; however, the B=200 moving-block bootstrap does not reject the fitted order-1 null, and edgewise FDR finds no localized discoveries. The empirical conclusion is calibrated caution, not proof of a latent neural driver.

The analysis combines surrogate-null calibration, VAR-null adequacy diagnostics, depth selection, edgewise localization, LPCMCI sensitivity checks, controls, mechanism ablations, power analysis, and nonlinear-CI sensitivity checks. These layers make the empirical claim calibrated rather than purely descriptive. The main limitation is attribution: latent common drive, lag-order misspecification, measurement error, nonstationarity, undersampling, nonlinear dynamics, and finite-sample testing error can all produce the same warning signal. A second limitation is calibration, because the v2a bootstrap is conditional on a stable but residual-autocorrelated VAR(1) null. A third limitation is cost, because the diagnostic reruns structure learning across depth grids and bootstrap replicates.

    \section*{Funding}
    This study was funded by the Zebrafish Neuroscience Interdisciplinary  Training Hub (ZENITH) program under Marie Curie Actions grant number \#813457.

    \bibliographystyle{unsrtnat5}
    \bibliography{references}

@book{spirtes2000causation,
  title={Causation, prediction, and search},
  author={Spirtes, Peter and Glymour, Clark N and Scheines, Richard},
  year={2000},
  publisher={MIT press}
}

@CONFERENCE{bica2020time,
  title={Time series deconfounder: Estimating treatment effects over time in the presence of hidden confounders},
  author={Bica, Ioana and Alaa, Ahmed and Van Der Schaar, Mihaela},
  booktitle={International Conference on Machine Learning},
  pages={884--895},
  year={2020},
  organization={PMLR}
}

@ARTICLE{louizos2017causal,
  title={Causal effect inference with deep latent-variable models},
  author={Louizos, Christos and Shalit, Uri and Mooij, Joris M and Sontag, David and Zemel, Richard and Welling, Max},
  journal={Advances in neural information processing systems},
  volume={30},
  year={2017}
}

@article{penning2023negative,
  author    = {Penning de Vries, Bas BL and Groenwold, Rolf HH},
  title     = {Negative controls: Concepts and caveats},
  journal   = {Statistical Methods in Medical Research},
  year      = {2023},
  volume    = {32},
  number    = {8},
  pages     = {1576--1587},
  doi       = {10.1177/09622802231181230}
}

@article{adedayo2025re,
  title={Re-examining Granger Causality from Causal Bayesian Networks Perspective},
  author={Adedayo, SA},
  journal={arXiv preprint arXiv:2501.02672},
  year={2025}
}

@article{karlsson2023detecting,
  author    = {Karlsson, Robin K. A. and Krijthe, Jesse H.},
  title     = {Detecting hidden confounding in observational data using multiple environments},
  journal   = {arXiv preprint arXiv:2205.13935},
  note      = {Presented at NeurIPS 2023},
  year      = {2023}
}

@article{benjamini1995,
  author  = {Benjamini, Yoav and Hochberg, Yosef},
  title   = {Controlling the False Discovery Rate: A Practical and Powerful Approach to Multiple Testing},
  journal = {Journal of the Royal Statistical Society Series B},
  volume  = {57},
  number  = {1},
  pages   = {289--300},
  year    = {1995},
  doi     = {10.1111/j.2517-6161.1995.tb02031.x}
}

@article{granger1969,
  author  = {Granger, Clive W. J.},
  title   = {Investigating Causal Relations by Econometric Models and Cross-Spectral Methods},
  journal = {Econometrica},
  volume  = {37},
  number  = {3},
  pages   = {424--438},
  year    = {1969},
  doi     = {10.2307/1912791}
}

@article{kunsch1989,
  author  = {K{\"u}nsch, Hans R.},
  title   = {The Jackknife and the Bootstrap for General Stationary Observations},
  journal = {Annals of Statistics},
  volume  = {17},
  number  = {3},
  pages   = {1217--1241},
  year    = {1989},
  doi     = {10.1214/aos/1176347265}
}

@article{politis1994,
  author  = {Politis, Dimitris N. and Romano, Joseph P.},
  title   = {The Stationary Bootstrap},
  journal = {Journal of the American Statistical Association},
  volume  = {89},
  number  = {428},
  pages   = {1303--1313},
  year    = {1994},
  doi     = {10.1080/01621459.1994.10476870}
}

@article{chen2024discovery,
  author  = {Chen, Li and Li, Chunlin and Shen, Xiaotong and Pan, Wei},
  title   = {Discovery and Inference of a Causal Network with Hidden Confounding},
  journal = {Journal of the American Statistical Association},
  volume  = {119},
  number  = {548},
  pages   = {2572--2584},
  year    = {2024},
  doi     = {10.1080/01621459.2023.2261658}
}

@article{colombo2012rfci,
  author  = {Colombo, Diego and Maathuis, Marloes H. and Kalisch, Markus and Richardson, Thomas S.},
  title   = {Learning High-Dimensional Directed Acyclic Graphs with Latent and Selection Variables},
  journal = {Annals of Statistics},
  volume  = {40},
  number  = {1},
  pages   = {294--321},
  year    = {2012},
  doi     = {10.1214/11-AOS940}
}

@article{damour2019multicause,
  author  = {D'Amour, Alexander},
  title   = {On Multi-Cause Causal Inference with Unobserved Confounding: Counterexamples, Impossibility, and Alternatives},
  journal = {arXiv preprint arXiv:1902.10286},
  year    = {2019},
  url     = {https://arxiv.org/abs/1902.10286}
}

@article{foygel2012halftrek,
  author  = {Foygel, Rina and Draisma, Jan and Drton, Mathias},
  title   = {Half-Trek Criterion for Generic Identifiability of Linear Structural Equation Models},
  journal = {Annals of Statistics},
  volume  = {40},
  number  = {3},
  pages   = {1682--1713},
  year    = {2012},
  doi     = {10.1214/12-AOS1012}
}

@inproceedings{entner2010fci,
  author    = {Entner, Dorothea and Hoyer, Patrik O.},
  title     = {On Causal Discovery from Time Series Data Using {FCI}},
  booktitle = {Proceedings of the Fifth European Workshop on Probabilistic Graphical Models},
  pages     = {121--128},
  year      = {2010}
}

@article{eichler2007granger,
  author  = {Eichler, Michael},
  title   = {Granger Causality and Path Diagrams for Multivariate Time Series},
  journal = {Journal of Econometrics},
  volume  = {137},
  number  = {2},
  pages   = {334--353},
  year    = {2007},
  doi     = {10.1016/j.jeconom.2006.06.016}
}

@article{eichler2012graphical,
  author  = {Eichler, Michael},
  title   = {Graphical Modelling of Multivariate Time Series},
  journal = {Probability Theory and Related Fields},
  volume  = {153},
  number  = {1--2},
  pages   = {233--268},
  year    = {2012},
  doi     = {10.1007/s00440-011-0345-8}
}

@article{eichler2013multiple,
  author  = {Eichler, Michael},
  title   = {Causal Inference with Multiple Time Series: Principles and Problems},
  journal = {Philosophical Transactions of the Royal Society A},
  volume  = {371},
  number  = {1997},
  pages   = {20110613},
  year    = {2013},
  doi     = {10.1098/rsta.2011.0613}
}

@inproceedings{gerhardus2020lpcmci,
  author    = {Gerhardus, Andreas and Runge, Jakob},
  title     = {High-Recall Causal Discovery for Autocorrelated Time Series with Latent Confounders},
  booktitle = {Advances in Neural Information Processing Systems},
  volume    = {33},
  pages     = {12615--12625},
  year      = {2020},
}

@article{hoyer2008hidden,
  author  = {Hoyer, Patrik O. and Shimizu, Shohei and Kerminen, Antti J. and Palviainen, Markus},
  title   = {Estimation of Causal Effects Using Linear Non-Gaussian Causal Models with Hidden Variables},
  journal = {International Journal of Approximate Reasoning},
  volume  = {49},
  number  = {2},
  pages   = {362--378},
  year    = {2008},
  doi     = {10.1016/j.ijar.2008.02.006}
}

@article{kuroki2014proxy,
  author  = {Kuroki, Manabu and Pearl, Judea},
  title   = {Measurement Bias and Effect Restoration in Causal Inference},
  journal = {Biometrika},
  volume  = {101},
  number  = {2},
  pages   = {423--437},
  year    = {2014},
  doi     = {10.1093/biomet/ast066}
}

@article{hyttinen2017subsampled,
  author  = {Hyttinen, Antti and Plis, Sergey and Jarvisalo, Matti and Eberhardt, Frederick and Danks, David},
  title   = {A Constraint Optimization Approach to Causal Discovery from Subsampled Time Series Data},
  journal = {International Journal of Approximate Reasoning},
  volume  = {90},
  pages   = {208--225},
  year    = {2017},
  doi     = {10.1016/j.ijar.2017.07.009}
}

@article{kummerfeld2016fofc,
  author  = {Kummerfeld, Erich and Ramsey, Joseph},
  title   = {Causal Clustering for 1-Factor Measurement Models},
  journal = {Proceedings of the 22nd ACM SIGKDD International Conference on Knowledge Discovery and Data Mining},
  pages   = {1655--1664},
  year    = {2016},
  doi     = {10.1145/2939672.2939838}
}

@article{lipsitch2010negative,
  author  = {Lipsitch, Marc and Tchetgen Tchetgen, Eric and Cohen, Ted},
  title   = {Negative Controls: A Tool for Detecting Confounding and Bias in Observational Studies},
  journal = {Epidemiology},
  volume  = {21},
  number  = {3},
  pages   = {383--388},
  year    = {2010},
  doi     = {10.1097/EDE.0b013e3181d61eeb}
}

@article{malinsky2018timeseries,
  author  = {Malinsky, Daniel and Spirtes, Peter},
  title   = {Causal Structure Learning from Multivariate Time Series in Settings with Unmeasured Confounding},
  journal = {Proceedings of Machine Learning Research},
  volume  = {92},
  pages   = {23--47},
  year    = {2018},
}

@article{miao2018proxy,
  author  = {Miao, Wang and Geng, Zhi and Tchetgen Tchetgen, Eric J.},
  title   = {Identifying Causal Effects with Proxy Variables of an Unmeasured Confounder},
  journal = {Biometrika},
  volume  = {105},
  number  = {4},
  pages   = {987--993},
  year    = {2018},
  doi     = {10.1093/biomet/asy038}
}

@inproceedings{ogarrio2016gfci,
  author    = {Ogarrio, Juan Miguel and Spirtes, Peter and Ramsey, Joe},
  title     = {A Hybrid Causal Search Algorithm for Latent Variable Models},
  booktitle = {Proceedings of the Eighth International Conference on Probabilistic Graphical Models},
  pages     = {368--379},
  year      = {2016},
  volume    = {52},
  series    = {Proceedings of Machine Learning Research},
  publisher = {PMLR},
}

@article{pearl1995diagrams,
  author  = {Pearl, Judea},
  title   = {Causal Diagrams for Empirical Research},
  journal = {Biometrika},
  volume  = {82},
  number  = {4},
  pages   = {669--688},
  year    = {1995},
  doi     = {10.1093/biomet/82.4.669}
}

@book{pearl2009causality,
  author    = {Pearl, Judea},
  title     = {Causality: Models, Reasoning, and Inference},
  edition   = {2},
  publisher = {Cambridge University Press},
  address   = {Cambridge},
  year      = {2009}
}

@book{peters2017elements,
  author    = {Peters, Jonas and Janzing, Dominik and Scholkopf, Bernhard},
  title     = {Elements of Causal Inference: Foundations and Learning Algorithms},
  publisher = {MIT Press},
  address   = {Cambridge, MA},
  year      = {2017}
}

@article{richardson2002ancestral,
  author  = {Richardson, Thomas and Spirtes, Peter},
  title   = {Ancestral Graph Markov Models},
  journal = {Annals of Statistics},
  volume  = {30},
  number  = {4},
  pages   = {962--1030},
  year    = {2002},
  doi     = {10.1214/aos/1031689015}
}

@article{runge2018reconstruction,
  author  = {Runge, Jakob},
  title   = {Causal Network Reconstruction from Time Series: From Theoretical Assumptions to Practical Estimation},
  journal = {Chaos: An Interdisciplinary Journal of Nonlinear Science},
  volume  = {28},
  number  = {7},
  pages   = {075310},
  year    = {2018},
  doi     = {10.1063/1.5025050}
}

@article{runge2019pcmci,
  author  = {Runge, Jakob and Nowack, Peer and Kretschmer, Marlene and Flaxman, Seth and Sejdinovic, Dino},
  title   = {Detecting and Quantifying Causal Associations in Large Nonlinear Time Series Datasets},
  journal = {Science Advances},
  volume  = {5},
  number  = {11},
  pages   = {eaau4996},
  year    = {2019},
  doi     = {10.1126/sciadv.aau4996}
}

@inproceedings{runge2020pcmcip,
  author    = {Runge, Jakob},
  title     = {Discovering Contemporaneous and Lagged Causal Relations in Autocorrelated Nonlinear Time Series Datasets},
  booktitle = {Proceedings of the 36th Conference on Uncertainty in Artificial Intelligence},
  pages     = {1388--1397},
  year      = {2020},
  volume    = {124},
  series    = {Proceedings of Machine Learning Research},
  publisher = {PMLR},
}

@inproceedings{gunther2023jpcmciplus,
  author    = {G{\"u}nther, Wiebke and Ninad, Urmi and Runge, Jakob},
  title     = {Causal Discovery for Time Series from Multiple Datasets with Latent Contexts},
  booktitle = {Proceedings of the Thirty-Ninth Conference on Uncertainty in Artificial Intelligence},
  pages     = {766--776},
  year      = {2023},
  editor    = {Evans, Robin J. and Shpitser, Ilya},
  volume    = {216},
  series    = {Proceedings of Machine Learning Research},
  month     = {31 Jul--04 Aug},
  publisher = {PMLR},
}

@article{runge2023review,
  author  = {Runge, Jakob and Gerhardus, Andreas and Varando, Gherardo and Eyring, Veronika and Camps-Valls, Gustau},
  title   = {Causal Inference for Time Series},
  journal = {Nature Reviews Earth \& Environment},
  volume  = {4},
  number  = {7},
  pages   = {487--505},
  year    = {2023},
  doi     = {10.1038/s43017-023-00431-y}
}

@article{shi2020negative,
  author  = {Shi, Xu and Miao, Wang and Tchetgen Tchetgen, Eric J.},
  title   = {A Selective Review of Negative Control Methods in Epidemiology},
  journal = {Current Epidemiology Reports},
  volume  = {7},
  number  = {4},
  pages   = {190--202},
  year    = {2020},
  doi     = {10.1007/s40471-020-00243-4}
}

@article{shi2020multiply,
  author  = {Shi, Xu and Miao, Wang and Nelson, Jennifer C. and Tchetgen Tchetgen, Eric J.},
  title   = {Multiply Robust Causal Inference with Double-Negative Control Adjustment for Categorical Unmeasured Confounding},
  journal = {Journal of the Royal Statistical Society: Series B},
  volume  = {82},
  number  = {2},
  pages   = {521--540},
  year    = {2020},
  doi     = {10.1111/rssb.12361}
}

@article{shimizu2006lingam,
  author  = {Shimizu, Shohei and Hoyer, Patrik O. and Hyvarinen, Aapo and Kerminen, Antti},
  title   = {A Linear Non-Gaussian Acyclic Model for Causal Discovery},
  journal = {Journal of Machine Learning Research},
  volume  = {7},
  number  = {72},
  pages   = {2003--2030},
  year    = {2006},
}

@article{shimizu2011directlingam,
  author  = {Shimizu, Shohei and Inazumi, Takanori and Sogawa, Yasuhiro and Hyvarinen, Aapo and Kawahara, Yoshinobu and Washio, Takashi and Hoyer, Patrik O. and Bollen, Kenneth},
  title   = {DirectLiNGAM: A Direct Method for Learning a Linear Non-Gaussian Structural Equation Model},
  journal = {Journal of Machine Learning Research},
  volume  = {12},
  number  = {33},
  pages   = {1225--1248},
  year    = {2011},
}

@article{silva2006latent,
  author  = {Silva, Ricardo and Scheines, Richard and Glymour, Clark and Spirtes, Peter},
  title   = {Learning the Structure of Linear Latent Variable Models},
  journal = {Journal of Machine Learning Research},
  volume  = {7},
  number  = {8},
  pages   = {191--246},
  year    = {2006},
}

@inproceedings{shpitser2006id,
  title={Identification of joint interventional distributions in recursive semi-Markovian causal models},
  author={Shpitser, Ilya and Pearl, Judea},
  booktitle={AAAI},
  pages={1219--1226},
  year={2006}
}

@article{shpitser2008hierarchy,
  author  = {Shpitser, Ilya and Pearl, Judea},
  title   = {Complete Identification Methods for the Causal Hierarchy},
  journal = {Journal of Machine Learning Research},
  volume  = {9},
  number  = {64},
  pages   = {1941--1979},
  year    = {2008},
}

@inproceedings{spirtes1995fci,
  author    = {Spirtes, Peter and Meek, Christopher and Richardson, Thomas S.},
  title     = {Causal Inference in the Presence of Latent Variables and Selection Bias},
  booktitle = {Proceedings of the Eleventh Conference on Uncertainty in Artificial Intelligence},
  pages     = {499--506},
  year      = {1995},
}

@article{tchetgen2020proximal,
  author  = {Tchetgen Tchetgen, Eric J. and Ying, Andrew and Cui, Yifan and Shi, Xu and Miao, Wang},
  title   = {An Introduction to Proximal Causal Learning},
  journal = {arXiv preprint arXiv:2009.10982},
  year    = {2020},
  doi     = {10.48550/arXiv.2009.10982},
}

@article{cui2023semiparametric,
  author  = {Cui, Yifan and Pu, Hongming and Shi, Xu and Miao, Wang and Tchetgen Tchetgen, Eric J.},
  title   = {Semiparametric Proximal Causal Inference},
  journal = {Journal of the American Statistical Association},
  volume  = {119},
  number  = {546},
  pages   = {1348--1359},
  year    = {2024},
  doi     = {10.1080/01621459.2023.2191817}
}

@inproceedings{tian2002testable,
  author    = {Tian, Jin and Pearl, Judea},
  title     = {On the Testable Implications of Causal Models with Hidden Variables},
  booktitle = {Proceedings of the Eighteenth Conference on Uncertainty in Artificial Intelligence},
  pages     = {519--527},
  year      = {2002},
}

@inproceedings{verma1990equivalence,
  author    = {Verma, Thomas S. and Pearl, Judea},
  title     = {Equivalence and Synthesis of Causal Models},
  booktitle = {Proceedings of the Sixth Conference on Uncertainty in Artificial Intelligence},
  pages     = {255--270},
  year      = {1990}
}

@article{sullivant2010trek,
  author  = {Sullivant, Seth and Talaska, Kelli and Draisma, Jan},
  title   = {Trek Separation for Gaussian Graphical Models},
  journal = {Annals of Statistics},
  volume  = {38},
  number  = {3},
  pages   = {1665--1685},
  year    = {2010},
  doi     = {10.1214/09-AOS760}
}

@article{wang2019deconfounder,
  author  = {Wang, Yixin and Blei, David M.},
  title   = {The Blessings of Multiple Causes},
  journal = {Journal of the American Statistical Association},
  volume  = {114},
  number  = {528},
  pages   = {1574--1596},
  year    = {2019},
  doi     = {10.1080/01621459.2019.1686987}
}

@article{zhang2008orientation,
  author  = {Zhang, Jiji},
  title   = {On the Completeness of Orientation Rules for Causal Discovery in the Presence of Latent Confounders and Selection Bias},
  journal = {Artificial Intelligence},
  volume  = {172},
  number  = {16--17},
  pages   = {1873--1896},
  year    = {2008},
  doi     = {10.1016/j.artint.2008.08.001}
}

@article{ahrens2013whole,
  author  = {Ahrens, Misha B. and Orger, Michael B. and Robson, Drew N. and Li, Jennifer M. and Keller, Philipp J.},
  title   = {Whole-Brain Functional Imaging at Cellular Resolution Using Light-Sheet Microscopy},
  journal = {Nature Methods},
  volume  = {10},
  number  = {5},
  pages   = {413--420},
  year    = {2013},
  doi     = {10.1038/nmeth.2434}
}

@article{barnett2009granger,
  author  = {Barnett, Lionel and Barrett, Adam B. and Seth, Anil K.},
  title   = {Granger Causality and Transfer Entropy Are Equivalent for Gaussian Variables},
  journal = {Physical Review Letters},
  volume  = {103},
  number  = {23},
  pages   = {238701},
  year    = {2009},
  doi     = {10.1103/PhysRevLett.103.238701}
}

@article{geweke1982linear,
  author  = {Geweke, John},
  title   = {Measurement of Linear Dependence and Feedback between Multiple Time Series},
  journal = {Journal of the American Statistical Association},
  volume  = {77},
  number  = {378},
  pages   = {304--313},
  year    = {1982},
  doi     = {10.1080/01621459.1982.10477803}
}

@article{geweke1984conditional,
  author  = {Geweke, John F.},
  title   = {Measures of Conditional Linear Dependence and Feedback between Time Series},
  journal = {Journal of the American Statistical Association},
  volume  = {79},
  number  = {388},
  pages   = {907--915},
  year    = {1984},
  doi     = {10.1080/01621459.1984.10477110}
}

@article{paninski2010state,
  author  = {Paninski, Liam and Ahmadian, Yashar and Ferreira, Daniel Gil and Koyama, Shinsuke and Rahnama Rad, Kamiar and Vidne, Michael and Vogelstein, Joshua and Wu, Wei},
  title   = {A New Look at State-Space Models for Neural Data},
  journal = {Journal of Computational Neuroscience},
  volume  = {29},
  number  = {1--2},
  pages   = {107--126},
  year    = {2010},
  doi     = {10.1007/s10827-009-0179-x}
}

@article{yu2009gpfa,
  author  = {Yu, Byron M. and Cunningham, John P. and Santhanam, Gopal and Ryu, Stephen I. and Shenoy, Krishna V. and Sahani, Maneesh},
  title   = {Gaussian-Process Factor Analysis for Low-Dimensional Single-Trial Analysis of Neural Population Activity},
  journal = {Journal of Neurophysiology},
  volume  = {102},
  number  = {1},
  pages   = {614--635},
  year    = {2009},
  doi     = {10.1152/jn.90941.2008}
}

@article{carbo2023mesencephalic,
  title={The mesencephalic locomotor region recruits V2a reticulospinal neurons to drive forward locomotion in larval zebrafish},
  author={Carbo-Tano, Martin and Lapoix, Mathilde and Jia, Xinyu and Thouvenin, Olivier and Pascucci, Marco and Auclair, Fran{\c{c}}ois and Quan, Feng B and Albadri, Shahad and Aguda, Vernie and Farouj, Younes and others},
  journal={Nature Neuroscience},
  volume={26},
  number={10},
  pages={1775--1790},
  year={2023},
  publisher={Nature Publishing Group US New York}
}

@article{ali2009markov,
  author  = {Ali, R. Ayesha and Richardson, Thomas S. and Spirtes, Peter},
  title   = {Markov Equivalence for Ancestral Graphs},
  journal = {Annals of Statistics},
  volume  = {37},
  number  = {5B},
  pages   = {2808--2837},
  year    = {2009},
  doi     = {10.1214/08-AOS626}
}

@article{glymour2019review,
  author  = {Glymour, Clark and Zhang, Kun and Spirtes, Peter},
  title   = {Review of Causal Discovery Methods Based on Graphical Models},
  journal = {Frontiers in Genetics},
  volume  = {10},
  pages   = {524},
  year    = {2019},
  doi     = {10.3389/fgene.2019.00524}
}

@article{assaad2022survey,
  author  = {Assaad, Charles K. and Devijver, Emilie and Gaussier, Eric},
  title   = {Survey and Evaluation of Causal Discovery Methods for Time Series},
  journal = {Journal of Artificial Intelligence Research},
  volume  = {73},
  pages   = {767--819},
  year    = {2022},
  doi     = {10.1613/jair.1.13428}
}

@inproceedings{eichler2010graphical,
  author    = {Eichler, Michael},
  title     = {Graphical Gaussian Modelling of Multivariate Time Series with Latent Variables},
  booktitle = {Proceedings of the Thirteenth International Conference on Artificial Intelligence and Statistics},
  series    = {Proceedings of Machine Learning Research},
  volume    = {9},
  pages     = {193--200},
  year      = {2010},
  publisher = {PMLR},
}

@article{kalisch2007pc,
  author  = {Kalisch, Markus and B{\"u}hlmann, Peter},
  title   = {Estimating High-Dimensional Directed Acyclic Graphs with the {PC}-Algorithm},
  journal = {Journal of Machine Learning Research},
  volume  = {8},
  number  = {22},
  pages   = {613--636},
  year    = {2007},
}

@article{colombo2014order,
  author  = {Colombo, Diego and Maathuis, Marloes H.},
  title   = {Order-Independent Constraint-Based Causal Structure Learning},
  journal = {Journal of Machine Learning Research},
  volume  = {15},
  number  = {116},
  pages   = {3921--3962},
  year    = {2014},
}

@article{sadeghi2022conditions,
  author  = {Sadeghi, Kayvan and Soo, Terry},
  title   = {Conditions and Assumptions for Constraint-Based Causal Structure Learning},
  journal = {Journal of Machine Learning Research},
  volume  = {23},
  number  = {109},
  pages   = {1--34},
  year    = {2022},
}

@article{richardson2023nested,
  author  = {Richardson, Thomas S. and Evans, Robin J. and Robins, James M. and Shpitser, Ilya},
  title   = {Nested Markov Properties for Acyclic Directed Mixed Graphs},
  journal = {Annals of Statistics},
  volume  = {51},
  number  = {1},
  year    = {2023},
  doi     = {10.1214/22-AOS2253}
}

@article{peters2016invariant,
  author  = {Peters, Jonas and B{\"u}hlmann, Peter and Meinshausen, Nicolai},
  title   = {Causal Inference by Using Invariant Prediction: Identification and Confidence Intervals},
  journal = {Journal of the Royal Statistical Society: Series B},
  volume  = {78},
  number  = {5},
  pages   = {947--1012},
  year    = {2016},
  doi     = {10.1111/rssb.12167}
}

@article{pfister2018invariant,
  author  = {Pfister, Niklas and B{\"u}hlmann, Peter and Peters, Jonas},
  title   = {Invariant Causal Prediction for Sequential Data},
  journal = {Journal of the American Statistical Association},
  volume  = {114},
  number  = {527},
  pages   = {1264--1276},
  year    = {2018},
  doi     = {10.1080/01621459.2018.1491403}
}

@article{tchetgen2024proximal,
  author  = {Tchetgen Tchetgen, Eric J. and Ying, Andrew and Cui, Yifan and Shi, Xu and Miao, Wang},
  title   = {An Introduction to Proximal Causal Inference},
  journal = {Statistical Science},
  volume  = {39},
  number  = {3},
  year    = {2024},
  doi     = {10.1214/23-STS911}
}

@article{chen2024cumulants,
  author  = {Chen, Wei and Huang, Zhiyi and Cai, Ruichu and Hao, Zhifeng and Zhang, Kun},
  title   = {Identification of Causal Structure with Latent Variables Based on Higher Order Cumulants},
  journal = {Proceedings of the AAAI Conference on Artificial Intelligence},
  volume  = {38},
  number  = {18},
  pages   = {20353--20361},
  year    = {2024},
  doi     = {10.1609/aaai.v38i18.30017}
}

@inproceedings{maeda2020rcd,
  author    = {Maeda, Takashi Nicholas and Shimizu, Shohei},
  title     = {{RCD}: Repetitive Causal Discovery of Linear Non-Gaussian Acyclic Models with Latent Confounders},
  booktitle = {Proceedings of the Twenty Third International Conference on Artificial Intelligence and Statistics},
  series    = {Proceedings of Machine Learning Research},
  volume    = {108},
  pages     = {735--745},
  year      = {2020},
  publisher = {PMLR},
}

@inproceedings{tramontano2024lingam,
  author    = {Tramontano, Daniele and Kivva, Yaroslav and Salehkaleybar, Saber and Drton, Mathias and Kiyavash, Negar},
  title     = {Causal Effect Identification in {LiNGAM} Models with Latent Confounders},
  booktitle = {Proceedings of the 41st International Conference on Machine Learning},
  series    = {Proceedings of Machine Learning Research},
  volume    = {235},
  pages     = {48468--48493},
  year      = {2024},
  publisher = {PMLR},
}

@inproceedings{liu2023strength,
  author    = {Liu, Yuhao and Cui, Chen and Waxman, Daniel and Butler, Kurt and Djuri{\'c}, Petar M.},
  title     = {Detecting Confounders in Multivariate Time Series Using Strength of Causation},
  booktitle = {2023 31st European Signal Processing Conference ({EUSIPCO})},
  pages     = {1400--1404},
  year      = {2023},
  publisher = {IEEE},
  doi       = {10.23919/EUSIPCO58844.2023.10289850}
}

@article{tank2022neural,
  author  = {Tank, Alex and Covert, Ian and Foti, Nicholas J. and Shojaie, Ali and Fox, Emily B.},
  title   = {Neural {G}ranger Causality},
  journal = {IEEE Transactions on Pattern Analysis and Machine Intelligence},
  volume  = {44},
  number  = {8},
  pages   = {4267--4279},
  year    = {2022},
  doi     = {10.1109/TPAMI.2021.3065601}
}

@inproceedings{zhou2024jrngc,
  author    = {Zhou, Wanqi and Bai, Shuanghao and Yu, Shujian and Zhao, Qibin and Chen, Badong},
  title     = {Jacobian Regularizer-Based Neural {G}ranger Causality},
  booktitle = {Proceedings of the 41st International Conference on Machine Learning},
  series    = {Proceedings of Machine Learning Research},
  volume    = {235},
  pages     = {61763--61782},
  year      = {2024},
  publisher = {PMLR},
}

@misc{zhang2025invargc,
  author        = {Zhang, Ziyi and Ren, Shaogang and Qian, Xiaoning and Duffield, Nick},
  title         = {{InvarGC}: Invariant {G}ranger Causality for Heterogeneous Interventional Time Series under Latent Confounding},
  year          = {2025},
  eprint        = {2510.19138},
  archivePrefix = {arXiv},
  primaryClass  = {cs.LG},
  doi           = {10.48550/arXiv.2510.19138}
}

    \clearpage
    \onecolumn
    \appendix
    \renewcommand{\thesection}{S\arabic{section}}
\renewcommand{\thefigure}{S\arabic{figure}}
\renewcommand{\thetable}{S\arabic{table}}
\setcounter{section}{0}
\setcounter{figure}{0}
\setcounter{table}{0}

\section{Supplementary Material}

The supplementary figures show supporting diagnostics that are important for auditability but too detailed for the main manuscript. They document the recovery-metric sanity checks, graph instability depth profiles, bootstrap calibration, controls, and mechanism ablations. These figures should be read as supporting evidence for the manuscript's calibrated-caution interpretation, not as additional claims of mechanism identification.

The tables provide the numerical audit trail behind that interpretation. Tables~\ref{tab:supp_simulation_qc}--\ref{tab:supp_mechanism_values} separate three questions: whether the controlled simulations behaved as intended, whether the biological recordings exceeded the fitted order-1 bootstrap null, and whether edge-level changes survived multiplicity correction. The figures then show the corresponding shapes of the depth curves and bootstrap distributions, so the reader can check that the tabulated summaries are not driven by a single hidden plotting choice.

\subsection{Numerical Audit Trail}

The retained graph instability analysis gives the cleanest simulation-level evidence. The order-1 negative control had \(T_{\mathrm{obs}}=0\) and zero cumulative instability in every retained repeat. The latent-common-driver condition produced positive maximum instability in 10/10 paired \acrshort{cgc} repeats and 8/10 paired \acrshort{fcgc} repeats. Mean \(T_{\mathrm{obs}}\) was 0.0333 for \acrshort{cgc} and 0.0189 for \acrshort{fcgc}; Holm-adjusted paired sign-flip tests for the primary \(T_{\mathrm{obs}}\) comparison gave \(p=0.0039\) and \(p=0.0078\), respectively. Table~\ref{tab:supp_simulation_qc} records the quality-control outcome that motivated retaining only the order-1 and latent-common-driver families.

\begin{table}[!htbp]
\centering
\small
\setlength{\tabcolsep}{4pt}
\renewcommand{\arraystretch}{1.18}
\caption{Graph instability numerical summary. Method trials combine the two optimized learners because both received the same saved input for each scenario and repeat.}
\label{tab:supp_simulation_qc}
\begin{tabular}{p{0.25\textwidth}p{0.17\textwidth}p{0.50\textwidth}}
\hline
Scenario family & QC status & Main numerical result \\
\hline
Order-1 unconfounded & 20/20 trials retained & \(T_{\mathrm{obs}}=0\) and cumulative instability \(=0\) in all retained repeats. \\
Latent common driver & 20/20 trials retained & Mean \(T_{\mathrm{obs}}\): 0.0333 for \acrshort{cgc}, 0.0189 for \acrshort{fcgc}; mean cumulative instability: 0.0867 and 0.0444. \\
Order-3 unconfounded & 14/20 trials passed; excluded wholesale & Three explosive repeats; 43,690 non-finite values across method trials; not interpreted. \\
Variable-lag unconfounded & 14/20 trials passed; excluded wholesale & Three explosive repeats; 16,426 non-finite values across method trials; not interpreted. \\
\hline
\end{tabular}
\end{table}

The main lesson from Table~\ref{tab:supp_simulation_qc} is that the retained simulation evidence is deliberately conservative. The clean order-1 system behaved as a true negative control, the latent-common-driver system produced measurable depth instability, and the numerically unstable higher-order families were removed from interpretation rather than used selectively.

The v2a-RSN calibration used \(B=200\) moving-block bootstrap replicates for each recording-method pair. The empirical \(p\)-value resolution is therefore \(1/(B+1)=0.00498\). Across all eight recording-method pairs, \(T_{\mathrm{obs}}\) ranged from 0.048 to 0.083, while the corresponding 95\% bootstrap critical values ranged from 0.117 to 0.175. No pair rejected the fitted order-1 null (Table~\ref{tab:supp_v2a_calibration}).

\begin{table}[!htbp]
\centering
\footnotesize
\setlength{\tabcolsep}{3pt}
\renewcommand{\arraystretch}{1.18}
\caption{Calibrated v2a-RSN global graph instability summary for 25 traces per recording, combining nine emitters and 16 receivers.}
\label{tab:supp_v2a_calibration}
\begin{tabular}{p{0.13\textwidth}p{0.09\textwidth}p{0.17\textwidth}p{0.18\textwidth}p{0.17\textwidth}p{0.09\textwidth}}
\hline
Method & Pairs & \(T_{\mathrm{obs}}\) range & 95\% critical range & Global \(p\)-values & Rejections \\
\hline
\acrshort{cgc} & 4 & 0.048--0.083 & 0.127--0.175 & 1.000 for all pairs & 0/4 \\
\acrshort{fcgc} & 4 & 0.048--0.072 & 0.117--0.153 & 0.990--1.000 & 0/4 \\
\hline
\end{tabular}
\end{table}

Table~\ref{tab:supp_v2a_calibration} shows why the biological result is framed as descriptive rather than confirmatory. The observed instability values are consistently below the bootstrap thresholds for both \acrshort{cgc} variants, so the calibrated analysis does not support rejecting the fitted order-1 null for any recording.

The bootstrap calibration is conditional on the fitted VAR(1) null. We therefore checked that null directly using information criteria up to lag 5, VAR(1) stability, multivariate residual whiteness at lag 10, and univariate Ljung--Box tests at lag 10. All four VAR(1) fits were stable, but residual whiteness was rejected in every recording. AIC and FPE selected higher orders for all recordings, BIC selected order 1 only for fish-1 and order 2 for the remaining recordings, and HQIC selected order 2 or 3. These diagnostics do not overturn the graph-instability non-rejection; they show that the surrogate null is a stable calibration device rather than a complete model of the calcium traces.

\begin{table}[!htbp]
\centering
\small
\setlength{\tabcolsep}{4pt}
\renewcommand{\arraystretch}{1.18}
\caption{VAR-null adequacy diagnostics for the 25-trace v2a-RSN calibration profile. Information criteria were selected over candidate orders 1--5. Portmanteau and Ljung--Box diagnostics used lag 10; \(<10^{-16}\) indicates numerical underflow to zero in the reported multivariate test.}
\label{tab:supp_var_null_diagnostics}
\begin{tabular}{p{0.11\textwidth}rrrrp{0.12\textwidth}p{0.14\textwidth}p{0.17\textwidth}}
\hline
Fish & AIC & BIC & HQIC & FPE & VAR(1) stable & Portmanteau \(p\) & Ljung--Box significant traces \\
\hline
fish-1 & 5 & 1 & 2 & 5 & yes & \(<10^{-16}\) & 23/25 \\
fish-2 & 5 & 2 & 3 & 5 & yes & \(<10^{-16}\) & 25/25 \\
fish-3 & 5 & 2 & 3 & 5 & yes & \(<10^{-16}\) & 25/25 \\
fish-4 & 4 & 2 & 3 & 4 & yes & \(<10^{-16}\) & 25/25 \\
\hline
\end{tabular}
\end{table}

Table~\ref{tab:supp_var_null_diagnostics} explains why the main text treats the biological calibration as cautious. The fitted order-1 null is dynamically stable, so it can generate surrogate trajectories for calibration, but residual autocorrelation remains. The empirical statement is therefore narrower than ``the data are order-1 Markov'': the observed graph-instability statistic is not extreme relative to this fitted bootstrap null.

Edge localization was summarized numerically rather than plotted as heatmaps. The aggregate table contains 1,207 edge rows: 616 for \acrshort{cgc} and 591 for \acrshort{fcgc}. Stable edges appeared across the depth grid, unstable edges changed at least once, lost edges disappeared after being present, and novel edges appeared only after additional history was included. These labels describe where the graph changed; they are not discoveries by themselves. All rows were calibrated against the same 200 bootstrap adjacency replicates used in the global analysis. The smallest unadjusted edgewise \(p\)-value was 0.00498, but no edge passed false-discovery-rate correction at \(q\leq0.05\); the minimum method-level \(q\)-value was 1.0, and the minimum recording-method \(q\)-value was 0.461.

\begin{table}[!htbp]
\centering
\small
\setlength{\tabcolsep}{4pt}
\renewcommand{\arraystretch}{1.18}
\caption{Edge-localization aggregate status for the calibrated 25-trace v2a-RSN analysis, with nine emitters and 16 receivers per recording. Edge status labels are descriptive; calibrated discoveries require FDR survival.}
\label{tab:supp_edge_localization}
\begin{tabular}{p{0.14\textwidth}rrrrrp{0.21\textwidth}}
\hline
Method & Rows & Stable & Unstable & Lost & Novel & Calibration outcome \\
\hline
\acrshort{cgc} & 616 & 288 & 241 & 81 & 6 & 200 bootstrap replicates per row; no FDR discovery. \\
\acrshort{fcgc} & 591 & 287 & 224 & 70 & 10 & 200 bootstrap replicates per row; no FDR discovery. \\
Combined & 1207 & 575 & 465 & 151 & 16 & Minimum method-level \(q=1.0\). \\
\hline
\end{tabular}
\end{table}

Table~\ref{tab:supp_edge_localization} separates graph movement from calibrated discovery. Many edges change status across conditioning depths, especially through instability and loss, but the bootstrap-calibrated FDR screen removes them as formal edge-level findings.

Control summaries and mechanism ablations support the intended direction of the diagnostic but also reinforce mechanism non-specificity. Synthetic null controls had mean \(T_{\mathrm{obs}}=0.0156\) for both order-1 and order-3 summaries, while synthetic latent-common-driver and hidden-node controls had means 0.0333 and 0.0300. In real-data perturbations, time shuffling, phase randomization, and circular shifts were small (\(T_{\mathrm{obs}}\leq0.0045\)), whereas block shuffling preserved temporal structure and remained close to the biological baseline (0.0196 versus 0.0200).

\begin{table}[!htbp]
\centering
\small
\setlength{\tabcolsep}{4pt}
\renewcommand{\arraystretch}{1.18}
\caption{Mechanism-ablation mean \(T_{\mathrm{obs}}\) over ten repeats. These values show that several mechanisms can produce depth sensitivity, so the diagnostic should not be interpreted as unique latent-confounder attribution.}
\label{tab:supp_mechanism_values}
\begin{tabular}{p{0.25\textwidth}p{0.22\textwidth}rr}
\hline
Ablation group & Scenario & \acrshort{cgc} & \acrshort{fcgc} \\
\hline
\multirow{3}{*}{Hidden state} & Latent confounder & 0.0333 & 0.0189 \\
 & Hidden nodes & 0.0300 & 0.0267 \\
 & Omitted lag order & 0.0244 & 0.0256 \\
\hline
\multirow{2}{*}{Observation artifact} & Measurement noise & 0.0056 & 0.0056 \\
 & Undersampling & 0.0133 & 0.0111 \\
\hline
\multirow{2}{*}{Nonstationarity} & Time-varying coefficients & 0.0100 & 0.0100 \\
 & Regime shift & 0.0167 & 0.0167 \\
\hline
\end{tabular}
\end{table}

Table~\ref{tab:supp_mechanism_values} shows that depth sensitivity is not unique to one biological mechanism. Latent-state manipulations give the largest values, but omitted lag order, undersampling, and nonstationarity can also increase \(T_{\mathrm{obs}}\); this supports using the diagnostic as a warning signal rather than as a stand-alone causal label.

Two additional sensitivity outputs delimit the current evidence. The nonlinear conditional-independence sensitivity analysis completed only the ParCorr branch, with \(D_2=0.10,D_3=0.00\) for the order-1 control and \(D_2=0.05,D_3=0.15\) for the latent-common-driver scenario; nonlinear-residual testing failed and GPDC/CMIknn were unavailable because optional dependencies were missing. The sample-size and power grid covered 18 method-setting rows and 180 runs; all rows had mean TPR and mean FPR equal to zero, so these outputs are reported as operating-regime limitations rather than positive robustness claims.

\subsection{Supplementary Figure Notes}

\begin{figure}[!htbp]
    \centering
    \begin{subfigure}[t]{0.48\textwidth}
        \centering
        \includegraphics[width=\linewidth]{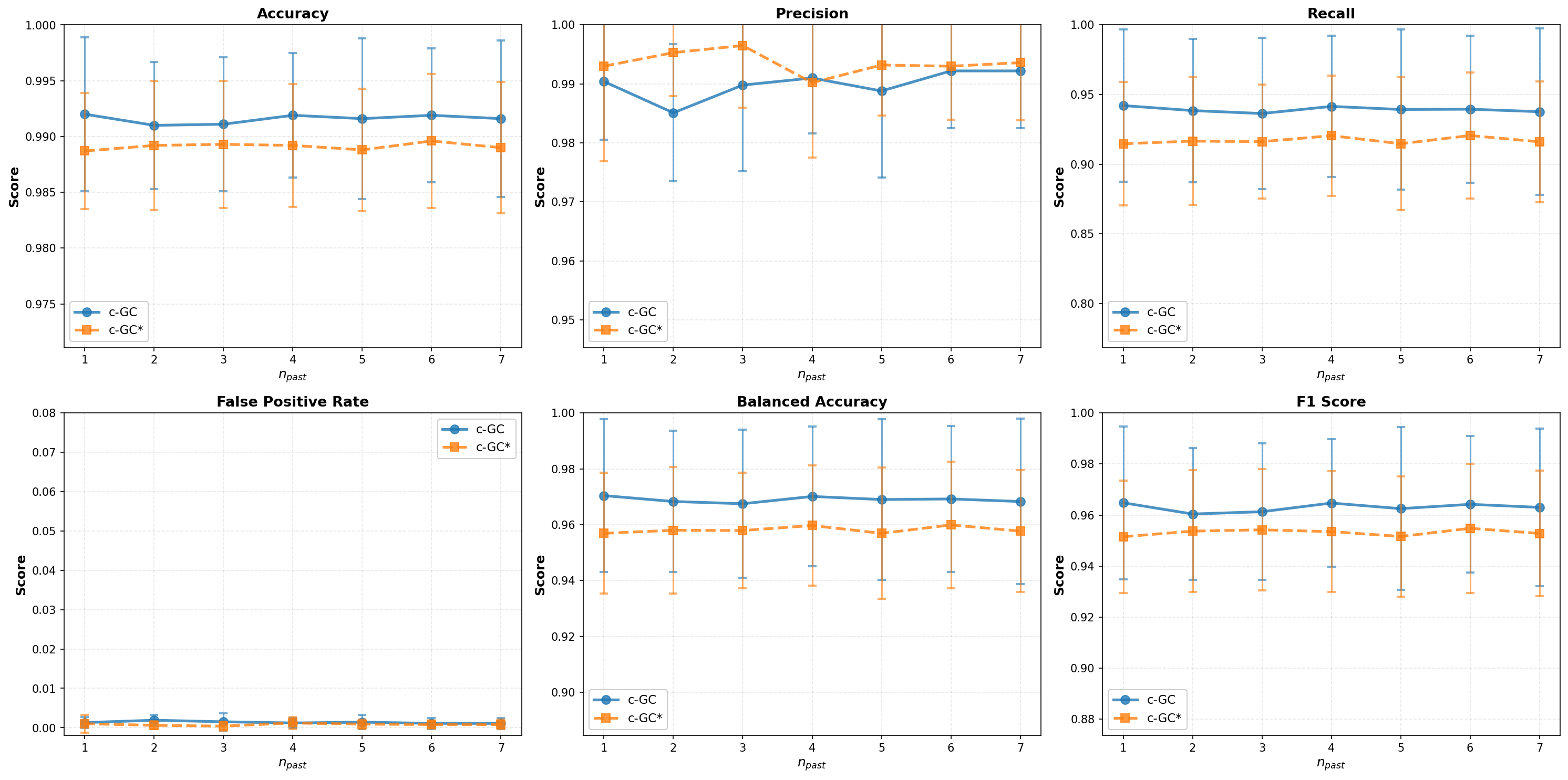}
        \caption{Single-lag Markovian.}
    \end{subfigure}
    \hfill
    \begin{subfigure}[t]{0.48\textwidth}
        \centering
        \includegraphics[width=\linewidth]{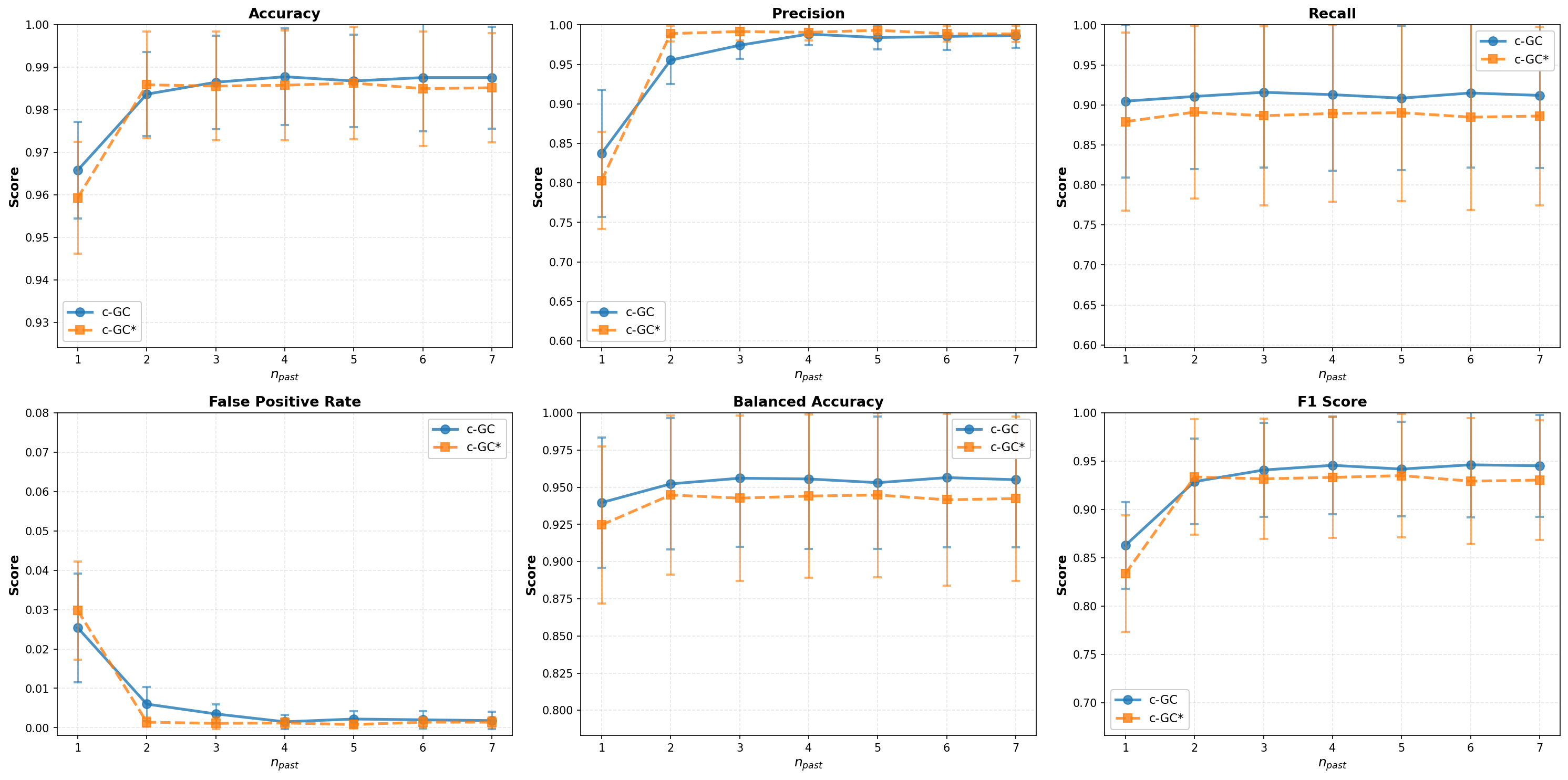}
        \caption{Single-lag hidden memory.}
    \end{subfigure}
    \vspace{0.45em}
    \begin{subfigure}[t]{0.48\textwidth}
        \centering
        \includegraphics[width=\linewidth]{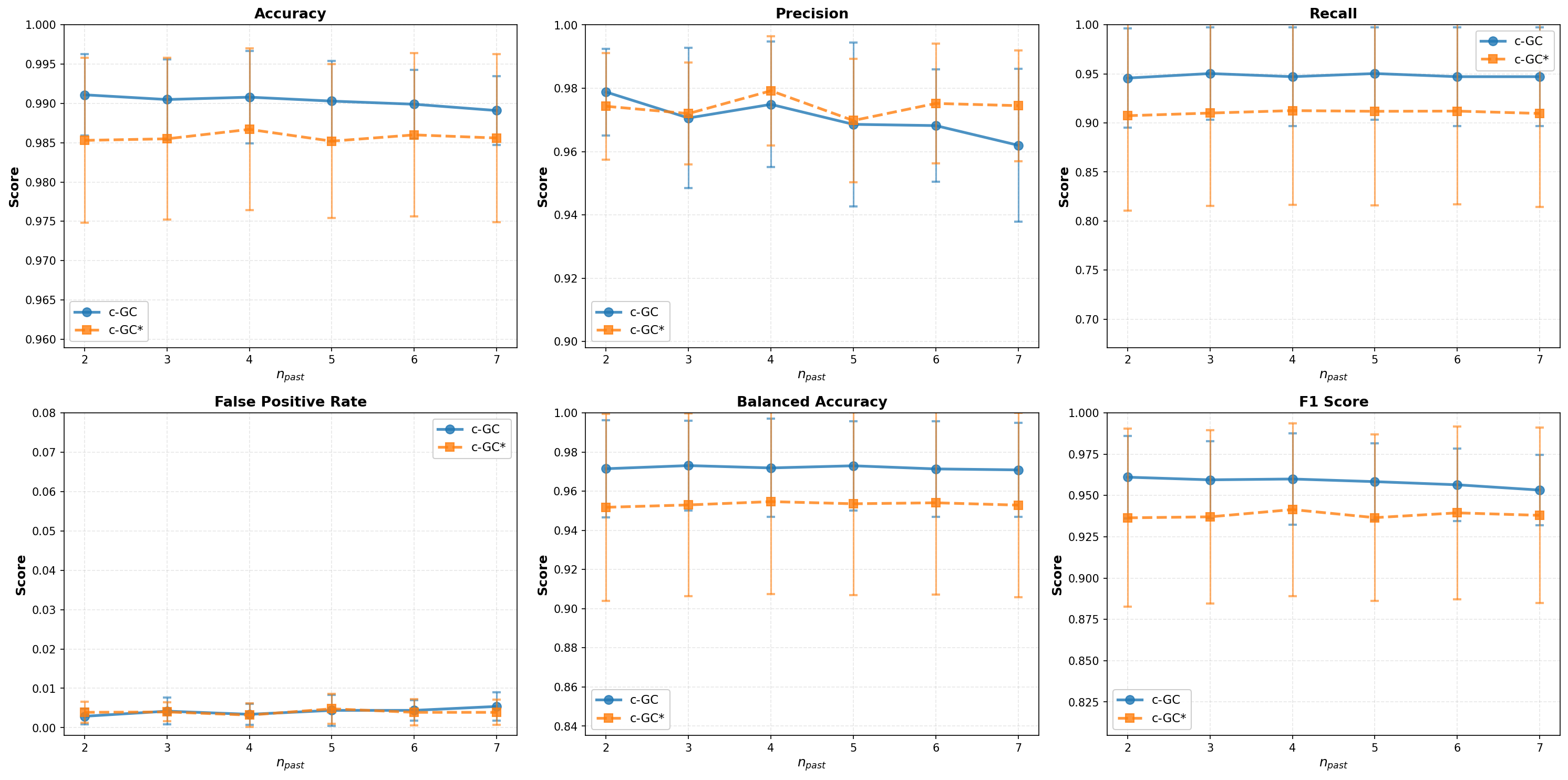}
        \caption{Multiple-lag Markovian.}
    \end{subfigure}
    \hfill
    \begin{subfigure}[t]{0.48\textwidth}
        \centering
        \includegraphics[width=\linewidth]{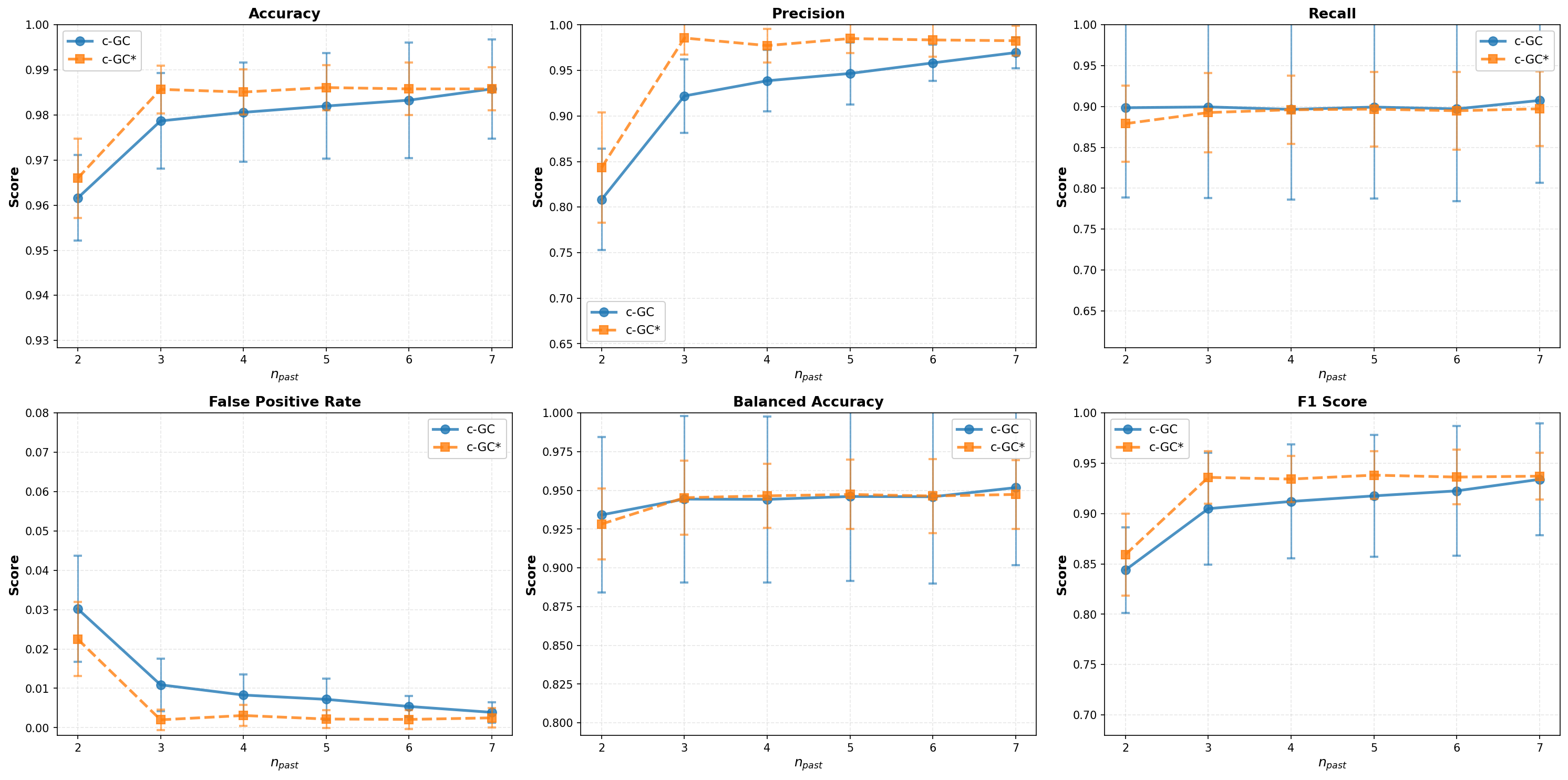}
        \caption{Multiple-lag hidden memory.}
    \end{subfigure}
    \caption{Supporting recovery-metric curves for the single- and multiple-lag simulations. Means and standard-deviation error bars are computed over 10 repeats. These plots check that the conventional graph-recovery metrics follow the expected direction, while the main text focuses on graph instability as the diagnostic statistic.}
    \label{fig:supp_recovery_curves}
\end{figure}

Figure~\ref{fig:supp_recovery_curves} contains the conventional recovery-metric curves that were moved out of the main text. These panels are useful sanity checks: the Markovian settings remain mostly stable, the hidden-memory settings improve false-positive-sensitive metrics as conditioning depth increases, and the heterogeneous multiple-lag setting is visibly harder for the shallower learner. The main manuscript therefore reports their conclusion in prose and reserves main-text figure space for the diagnostic statistic.

\begin{figure}[!htbp]
    \centering
    \includegraphics[width=0.74\textwidth]{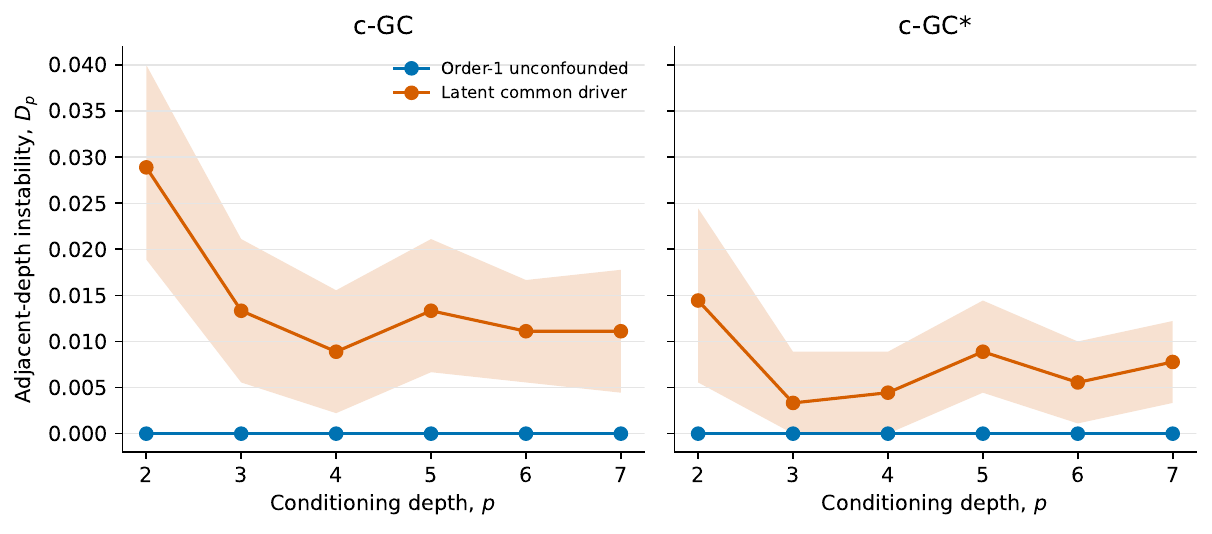}
    \caption{Adjacent-depth graph instability profiles. The clean order-1 process remains invariant across the depth grid, whereas the AR(1) latent-common-driver condition produces nonzero depth transitions under both \acrshort{cgc} variants.}
    \label{fig:supp_depth_profiles}
\end{figure}

Figure~\ref{fig:supp_depth_profiles} expands the main-text graph instability analysis by showing the full adjacent-depth profiles instead of only the scalar summaries. The important visual contrast is qualitative: the retained order-1 control remains flat, while the latent-common-driver runs show depth-sensitive departures under both optimized learners.

\begin{figure}[!htbp]
    \centering
    \includegraphics[width=0.78\textwidth]{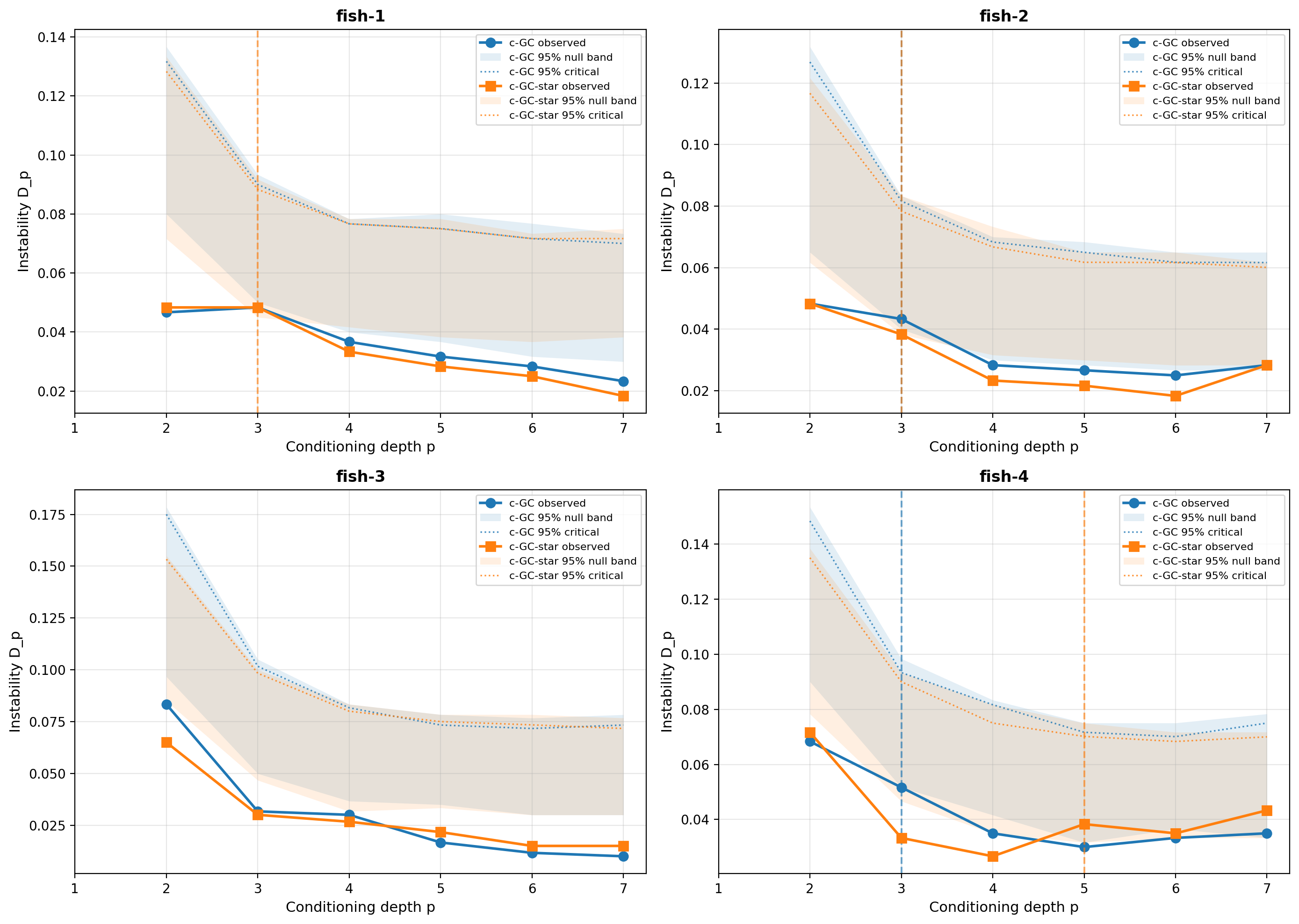}
    \caption{Calibrated v2a-RSN depth-selection curves for 25 traces per recording, combining nine emitters and 16 receivers. Observed graph instability trajectories are shown against the \(B=200\) moving-block bootstrap null bands and critical curves for each recording and method. The observed trajectories remain inside the fitted order-1 null envelope.}
    \label{fig:supp_v2a_calibrated_depth_selection}
\end{figure}

\begin{figure}[!htbp]
    \centering
    \begin{subfigure}[t]{0.48\textwidth}
        \centering
        \includegraphics[width=\linewidth]{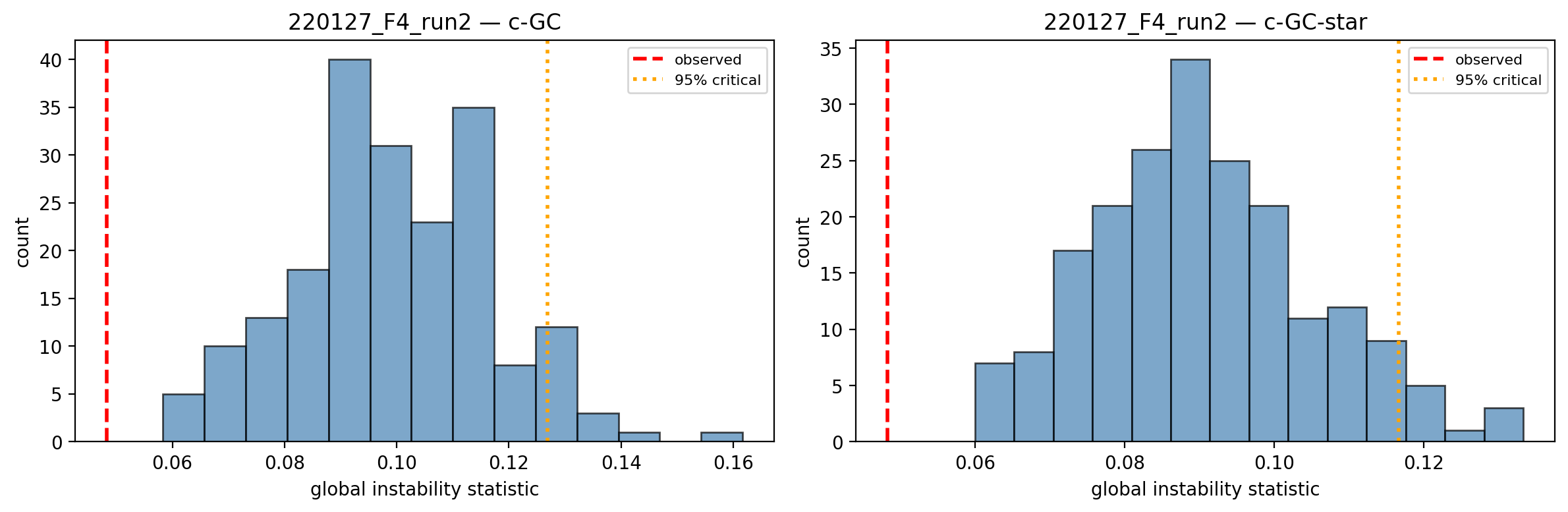}
        \caption{fish-2.}
    \end{subfigure}
    \hfill
    \begin{subfigure}[t]{0.48\textwidth}
        \centering
        \includegraphics[width=\linewidth]{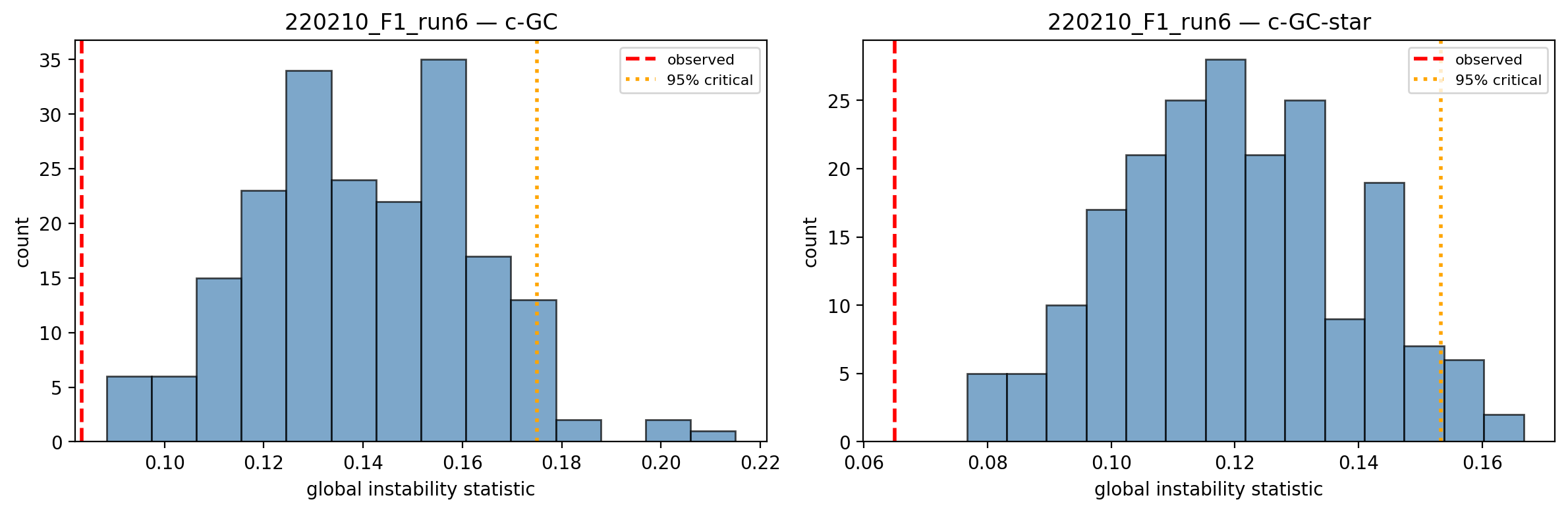}
        \caption{fish-3.}
    \end{subfigure}
    \vspace{0.45em}
    \begin{subfigure}[t]{0.48\textwidth}
        \centering
        \includegraphics[width=\linewidth]{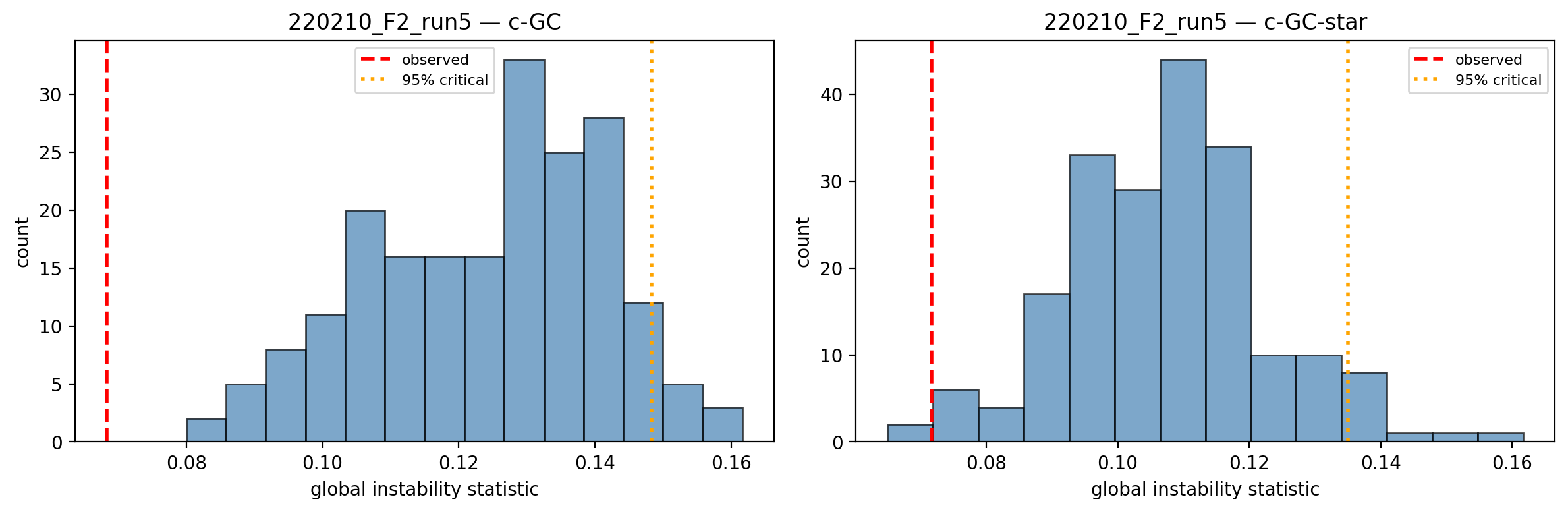}
        \caption{fish-4.}
    \end{subfigure}
    \caption{Additional bootstrap global-statistic distributions for the v2a-RSN recordings. Figure~\ref{fig:v2a-bootstrap-example} shows fish-1 in the main text; these panels show fish-2 through fish-4 and the same calibrated non-rejection pattern, with observed \(T_{\mathrm{obs}}\) values inside the \(B=200\) moving-block bootstrap null distributions.}
    \label{fig:supp_v2a_bootstrap_remaining}
\end{figure}

Figure~\ref{fig:supp_v2a_calibrated_depth_selection} shows the calibrated depth-selection curves. Figure~\ref{fig:v2a-bootstrap-example} in the main text gives fish-1 as the representative bootstrap calibration, while Figure~\ref{fig:supp_v2a_bootstrap_remaining} shows the remaining recordings.

\begin{figure}[!htbp]
    \centering
    \includegraphics[width=0.84\textwidth]{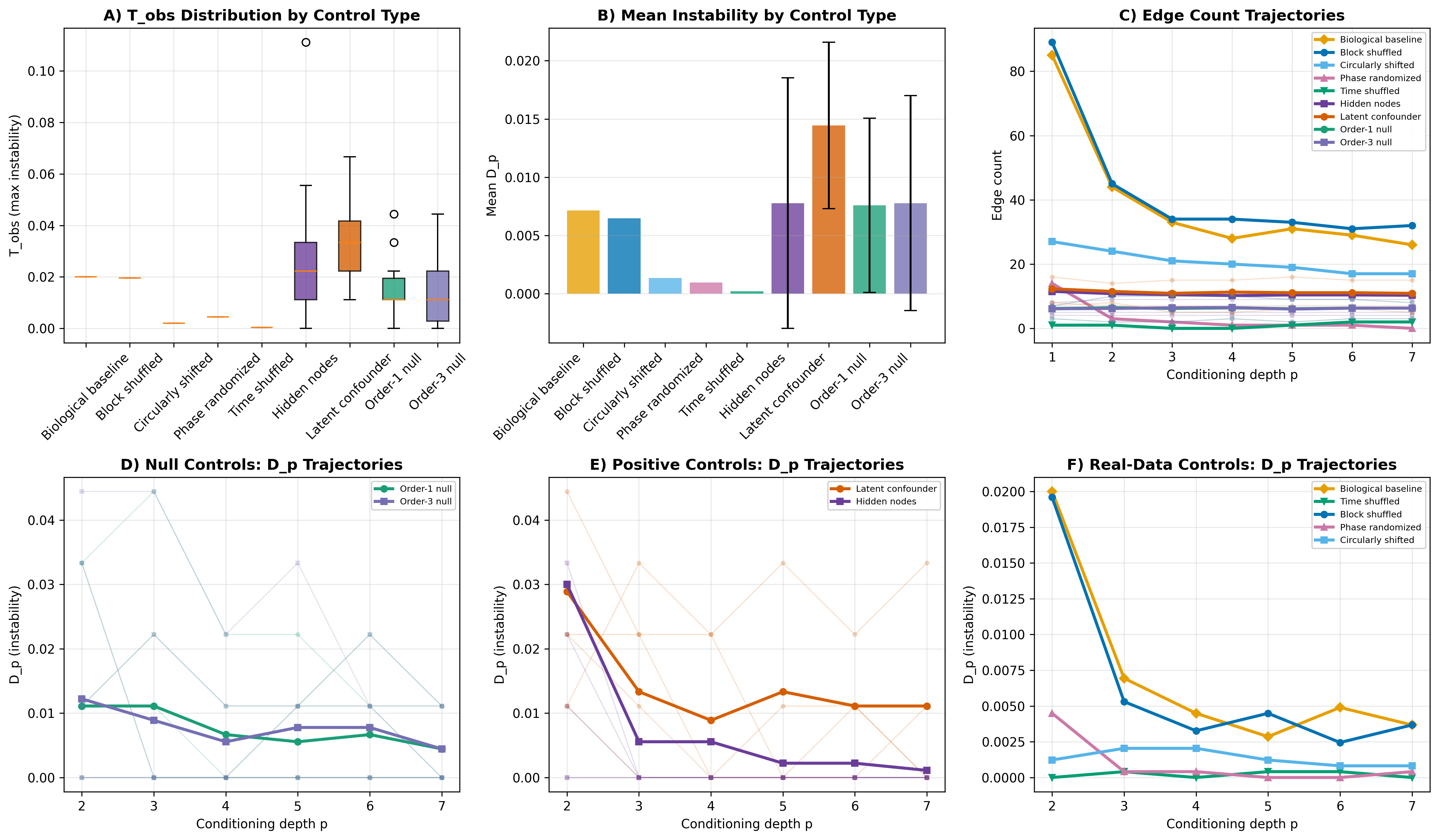}
    \caption{Positive and negative control summaries. Null and perturbation controls bound the expected direction of the diagnostic: hidden-state controls produce larger graph instability summaries than most shuffled or phase-randomized controls, while block shuffling retains temporal structure and remains closer to the biological baseline.}
    \label{fig:supp_controls}
\end{figure}

\begin{figure}[!htbp]
    \centering
    \begin{subfigure}[t]{0.82\textwidth}
        \centering
        \includegraphics[width=\linewidth]{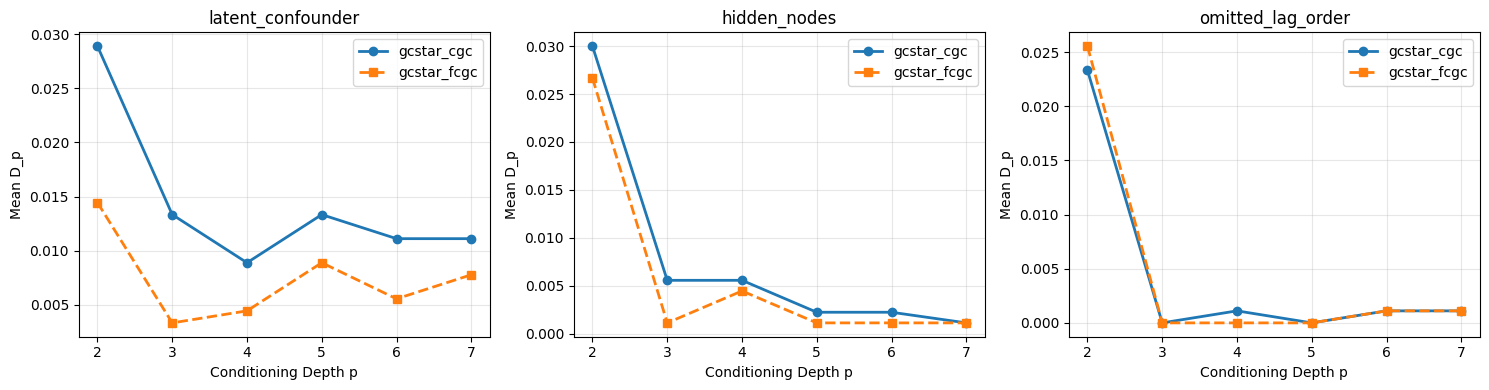}
        \caption{Hidden-state mechanisms.}
    \end{subfigure}
    \vspace{0.35em}
    \begin{subfigure}[t]{0.82\textwidth}
        \centering
        \includegraphics[width=\linewidth]{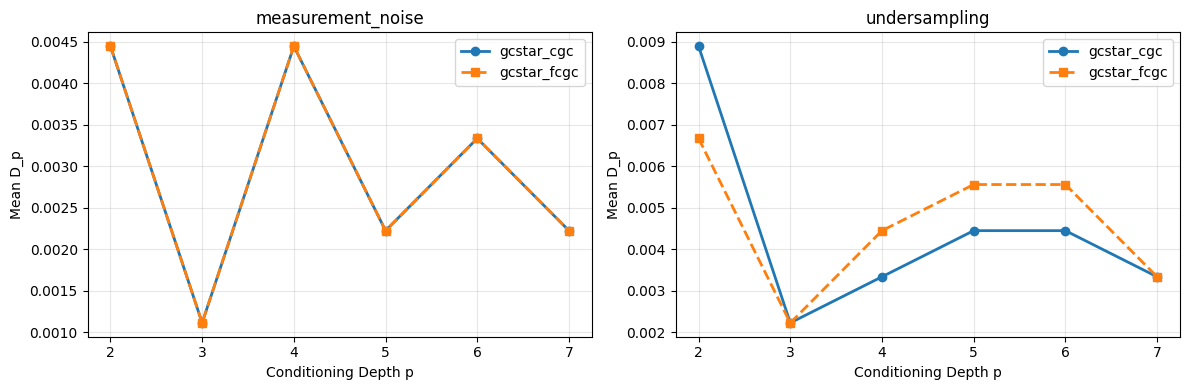}
        \caption{Observation-artifact mechanisms.}
    \end{subfigure}
    \vspace{0.35em}
    \begin{subfigure}[t]{0.82\textwidth}
        \centering
        \includegraphics[width=\linewidth]{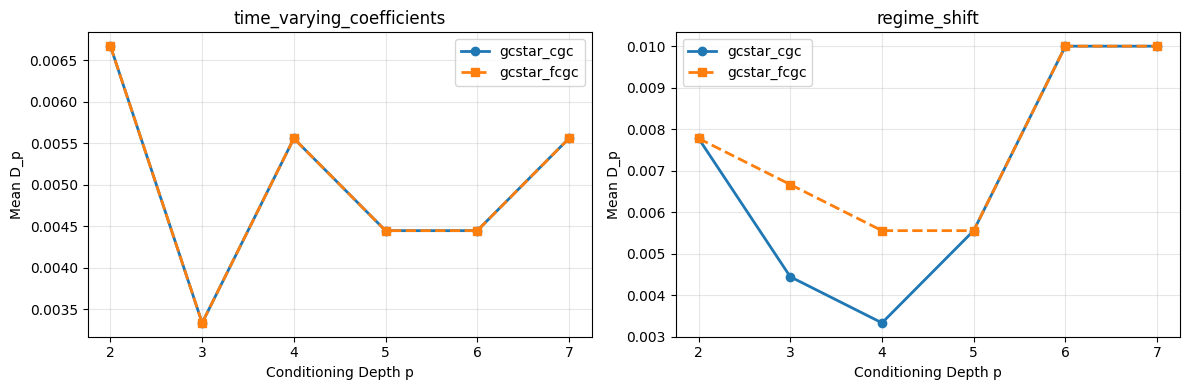}
        \caption{Nonstationarity mechanisms.}
    \end{subfigure}
    \caption{Mechanism-ablation depth profiles. The ablations support the mechanism-agnostic interpretation: depth sensitivity can arise from latent state, omitted lag order, observation artifacts, or nonstationarity, so the diagnostic should be interpreted as a hidden-memory warning rather than unique mechanism attribution.}
    \label{fig:supp_mechanism_ablations}
\end{figure}

Figures~\ref{fig:supp_controls} and \ref{fig:supp_mechanism_ablations} document the guardrails around interpretation. The control and ablation figures show that depth sensitivity is a warning sign for hidden memory or violated modeling assumptions, not a unique label for latent confounding. Edge localization is reported in Table~\ref{tab:supp_edge_localization}, where descriptive edge changes are separated from the stricter FDR outcome.

Supplementary Figure~\ref{fig:supp_recovery_curves} gives the expected sanity-check pattern. Markovian systems remain comparatively flat across depth, while hidden-memory conditions improve most clearly in false-positive-sensitive metrics; the multiple-lag setting is less clean because true predictive structure is distributed across delays.

Supplementary Figure~\ref{fig:supp_depth_profiles} shows the full trajectory behind the scalar graph-instability summary. The clean process stays flat, whereas the latent-driver process creates visible adjacent-depth movement, confirming that the main-text statistic is not hiding a single anomalous transition.

Supplementary Figure~\ref{fig:supp_v2a_calibrated_depth_selection} explains the calibrated caution in the real data. The solid marker curves are the observed adjacent-depth graph-instability values \(D_p\) for \acrshort{cgc} and \acrshort{fcgc}. The pale shaded regions are the 95\% moving-block bootstrap null bands expected under the fitted order-1 model; they function as an error envelope for the null trajectory rather than as uncertainty around the observed curve. The dotted curves are the pointwise 95\% critical values, and the vertical dashed lines mark the depth selected by the bootstrap-band rule when such a depth is available. In practical terms, a curve that remains below the dotted critical line and inside the shaded null band is not unusually unstable relative to the fitted order-1 null. Several recordings show an early drop after adding one extra past state, but the observed trajectories stay within the bootstrap envelope, so the figure supports descriptive stabilization rather than calibrated rejection.

Supplementary Figure~\ref{fig:supp_v2a_bootstrap_remaining} shows that the fish-1 example in the main text is representative rather than exceptional. Each histogram is the bootstrap null distribution of the global statistic \(T_{\mathrm{obs}}\), generated from \(B=200\) moving-block surrogate time series under the fitted order-1 model for that recording. The red dashed line is the observed global statistic from the real recording, and the orange dotted line is the 95\% bootstrap critical value. For fish-2 through fish-4, the observed statistic lies inside the bootstrap null distribution and below the 95\% critical value. These panels are therefore not statistically significant at the 5\% global level; they match Table~\ref{tab:supp_v2a_calibration}, where all shown pairs have \(p\geq0.990\) and no global rejection.

Supplementary Figure~\ref{fig:supp_controls} places the biological curves between deliberately disrupted and deliberately structured controls. Panels A and B summarize the global statistic and mean adjacent-depth instability by control type. The biological baseline is larger than time-shuffled, phase-randomized, and circular-shifted real-data controls, which largely remove or scramble temporal structure. The block-shuffled control remains close to the biological baseline because it preserves local temporal dependence within blocks, so it is a stricter negative control than complete shuffling. Panels C, D, E, and F show the underlying depth trajectories: edge counts and \(D_p\) curves change little for most null controls, while positive controls with latent common drive or hidden nodes produce larger and more variable instability. This figure is a diagnostic control check rather than a formal hypothesis test; its role is to show that the statistic responds to preserved hidden or temporal structure and is suppressed by controls that destroy that structure.

Supplementary Figure~\ref{fig:supp_mechanism_ablations} is the main guardrail against overinterpretation. The hidden-state panel shows the largest responses: latent confounding, hidden nodes, and omitted lag order all create depth-dependent graph movement. The observation-artifact panel shows smaller but nonzero responses under measurement noise and undersampling, meaning that acquisition or sampling artifacts can also make a shallow graph appear unstable. The nonstationarity panel shows intermediate responses for time-varying coefficients and regime shifts. Together with Table~\ref{tab:supp_mechanism_values}, the figure shows that depth sensitivity is mechanism-sensitive but not mechanism-specific: a large \(D_p\) trajectory flags inadequate observed state or violated modeling assumptions, but it does not by itself distinguish latent confounding from omitted lags, artifacts, or nonstationarity.

\end{document}